\documentclass{article}

\usepackage{arxiv}

\usepackage[utf8]{inputenc} 
\usepackage[T1]{fontenc}    
\usepackage[hidelinks]{hyperref}       
\usepackage{url}            
\usepackage{booktabs}       
\usepackage{amsfonts}       
\usepackage{nicefrac}       
\usepackage{microtype}      
\usepackage{graphicx}
\usepackage{doi}
\usepackage{lscape}
\usepackage[table]{xcolor}
\usepackage{arydshln}
\usepackage{array}
\usepackage{lmodern} 
\usepackage{caption}
\usepackage{bbm}
\usepackage{amsmath}
\usepackage{float} 
\usepackage[authoryear,round]{natbib}
\DeclareCaptionFont{customsize}{\fontsize{8pt}{11pt}\selectfont}
\definecolor{cdark1}{RGB}{201, 218, 238} 
\definecolor{cdark2}{RGB}{216, 223, 221} 
\definecolor{clight1}{RGB}{236, 242, 249} 
\definecolor{clight2}{RGB}{241, 244, 243} 

\newcommand{\boldhline}{
    \noalign{\global\arrayrulewidth=1.2pt}
    \hline
    \noalign{\global\arrayrulewidth=0.4pt}
}

\title{Forests of Uncertaint(r)ees: \\Using tree-based ensembles to estimate probability distributions of future conflict}

\date{March 12, 2026}	

\author{\large \href{https://orcid.org/0009-0003-6284-5017}{Daniel Mittermaier}$^{\dag}$\thanks{daniel.mittermaier@unibw.de} , 
	\href{https://orcid.org/0009-0009-6478-432X}{Tobias Bohne}$^{\dag}$,
	\href{https://orcid.org/0000-0003-3575-7620}{Martin Hofer},
    \href{https://orcid.org/0000-0002-9564-7880}{Daniel Racek}$^{\dag}$\\[1ex]
\normalfont\normalsize
  $^{\dag}$Center for Crisis Early Warning (CCEW), University of the Bundeswehr Munich
}

\renewcommand{\headeright}{}
\renewcommand{\undertitle}{}
\renewcommand{\shorttitle}{Forests of Uncertaint(r)ees}

\hypersetup{
pdftitle={Forest of Uncertaint(r)ees},
pdfsubject={stats.AP, cs.LG},
pdfauthor={Daniel Mittermaier, Tobias Bohne, Martin Hofer},
pdfkeywords={Conflict forecasting, Uncertainty quantification, Machine learning, Armed conflict},
}

\begin{document}
\maketitle
\setcounter{footnote}{0}
\begin{abstract}
Predictions of fatalities from violent conflict on the PRIO-GRID-month (\emph{pgm}) level are characterized by high levels of uncertainty, limiting their usefulness in practical applications. We discuss the two main sources of uncertainty for this prediction task, the nature of violent conflict and data limitations, embedding conflict prediction in the wider literature on uncertainty quantification in machine learning. Based on this, we develop a strategy to quantify uncertainty in conflict forecasting, shifting from traditional point predictions to full predictive distributions. Our approach combines multiple tree-based classifiers and distributional regressors in a custom AutoML setup, estimating distributions for each \emph{pgm} individually. We also test the integration of regional models in spatial ensembles as a potential avenue to reduce uncertainty by lowering data requirements and accounting for systematic differences between conflict contexts. The models are able to consistently outperform a suite of benchmarks derived from conflict history in predictions up to one year in advance. Marginal differences in model-wide metrics emphasize the need to understand their behavior for a given prediction problem, in this case characterized by extremely high zero-inflatedness. Adressing this, we compliment our evaluation with a simulation experiment, which demonstrates that our models reflect meaningful performance improvements, which can be traced back to conflict-affected regions. Lastly, we show that the integration of regional models does not decrease performance, opening avenues to integrate additional data sources in the future.
\end{abstract}


\section{Introduction}
\label{sec:introduction}
With the advent of fine-grained conflict data and the increasing availability of computational resources over the last decade and a half, significant efforts have been made to design conflict forecasting systems to warn policy-makers of emerging dangers and enable preventative action. These systems generally provide probability estimates for conflict occurrence or point estimates of conflict-related fatalities (see \citet{rod_review_2024} for a comparison of existing systems). However, \citet{hegre_202324_2025} highlight two key problems limiting the utility of both types of point predictions: First, they only reflect the most likely outcome. Second, they do not provide any confidence estimates for the predicted value. Importantly, both issues are closely linked, as the latter becomes more important the less likely the underlying scenario is. Their proposed solution is to estimate a predictive distribution, which allows for the quantification of uncertainties.

The sources of uncertainty in conflict prediction can be broken down into uncertainties related to the nature of conflict, and uncertainties related to the (lack of) data used in conflict prediction. The former includes the complexity of conflict and related challenges of translating explanation to prediction \citep{ward_perils_2010}, its relative rarity \citep{mueller_hard_2022}, its changing nature \citep{bowlsby_future_2020, hegre_can_2021}, and resulting doubts regarding the systematicity and therefore predictability of conflict \citep{chadefaux_conflict_2017}. The latter includes various data-related challenges in the field, such as concerns over quality, availability, and low resolution of data on potential predictors \citep[e.g.][]{cederman_predicting_2017, mueller_reading_2018}, and a number of biases in conflict event datasets, many related to its reliance on news media \citep{weidmann_accuracy_2015, althaus_total_2022, bazzi_promise_2022, oberg_measurement_2025, raleigh_political_2023}. Integrating this into a much larger body of research on uncertainty quantification (UQ) in machine learning (ML), these issues can also be viewed in terms of aleatoric and epistemic uncertainty, with the former resulting from inherently stochastic processes and therefore irreducible, and the latter related to data selection and the modeling process \citep{hullermeier_aleatoric_2021, gruber_sources_2025}. While there is room for debate about how much of the apparently stochastic nature of conflict \citep{chadefaux_conflict_2017} is in fact the product of data issues, from a statistical perspective the various methods to quantify uncertainty can be loosely summarized as estimating the predictive distribution for a given observation \citep{gruber_sources_2025}. This aligns with the approach proposed by \citet{hegre_202324_2025} for conflict forecasts.

We contribute to the understanding of uncertainty in conflict prediction by designing and evaluating a ML approach including UQ on the PRIO-GRID-month level \citep[\emph{pgm},][]{tollefsen_prio-grid_2012}, combining various tree-based models. We specifically incorporate two key sources of data-related uncertainty: First, we rely on algorithms which are able to estimate distributions natively and for each \emph{pgm} individually, accounting for local differences in data generation processes and conflict mechanisms. Second, we employ probabilistic hurdle ensembles to account for systematic differences between the recording of events and the recording of fatalities \citep{hegre_forecasting_2022, lacina_explaining_2006}. We select combinations of either Random Forest \citep{breiman_random_2001} or XGBoost \citep{chen_xgboost_2016} classifiers with Distributional Random Forest \citep{cevid_distributional_2022}, Quantile Regression Forest \citep{meinshausen_quantile_2006}, or NGBoost \citep{duan_ngboost_2019} regression models through a custom AutoML\footnote{AutoML refers to the automation of some or all components of machine learning, such as model or hyperparameter selection. For more information, see e.g. \citet{hutter_automated_2019}.} setup. Additionally, we explore an avenue to address data availability issues and regionally varying conflict mechanisms by integrating multiple regional models with limited geographical scope in a spatial ensemble. If successeful, data coverage requirements for individual models are reduced to the corresponding region, thus allowing for the incorporation of region-specific datasets. This yields three model specifications: a ``global''-only model covering the whole area of interest, a ``local''only model combining all regional models, and a global-local model based on the best-performing individual components of each.

Our evaluation shows that all three approaches outperform benchmarks across three distributional metrics with only very few exceptions. The global model and the global-local model score very similarly, while the local model performs only slightly worse. However, in absolute terms the differences in scores are marginal, prompting further investigation. Since the zero-inflatedness of our target greatly reduces the informative range of our scores, we design an experiment with simulated data to examine the impact of varying accuracy and noise in the predictions on our three metrics. The results show that the small differences in absolute scores observed likely reflect sizable differences in prediction quality for this problem. We complement the model-wide evaluation with a ranking-based approach, comparing the mean ranks based on performance across all countries. While this approach disadvantages our model compared to the benchmarks for countries without violence, when we restrict the evaluation to country-year instances with any observed violence – arguably the cases in which predictive performance is most relevant – our models clearly outperform the benchmarks.

\section{Sources of Uncertainty in Conflict Prediction}
\label{sec:theory}

Upon closer inspection, predicting violence with reasonable certainty is a daunting task. The many challenges can broadly be grouped into two categories: the nature of conflict and its determinants, and the characteristics, biases, and availability of the data used. The uncertainty resulting from these challenges means that point predictions, i.e. predictions without uncertainty estimates,  often struggle to beat even simple heuristics on common evaluation metrics \citep{vesco_united_2022}.

Decades of research into the causes of armed conflict has identified a number of contributing factors from opportunity and feasibility to motivation and grievances \citep[e.g.][]{fearon_ethnicity_2003, collier_beyond_2008, collier_greed_2004, blattman_civil_2010}, with differences between individuals and social groups \citep{humphreys_who_2008, ostby_polarization_2008, cederman_inequality_2013}. However, this does not automatically mean these findings translate into reliable predictions for the future \citep{ward_perils_2010, chadefaux_endogenous_2025}.  While they are known to increase the risk of conflict, no combination of these factors can be considered a sufficient condition for conflict, with very similar contexts resulting in widely different outcomes in practice. The complex nature of conflict together with the unpredictability of actors involved has even led some to question to what extent conflict is even systematic enough to be predicted \citep{chadefaux_conflict_2017}. The challenge is exacerbated by the relative rarity of conflict, especially at  the \emph{pgm}-level,\footnote{See \ref{sec:methodsdata}: From 1990 to 2017, UCDP records violence in less than 0.4\% of all \emph{pgms}.} which means in practice there are very few instances of conflict that can be used to derive very complex patterns from \citep{mueller_hard_2022, hegre_202324_2025}. Uncertainty is further increased by continuously evolving situations on the ground and larger geopolitical shifts, which can affect underlying risk patterns and limit the validity of historic insights or change the conditions that might lead to violence \citep{cederman_predicting_2017, chadefaux_conflict_2017, bowlsby_future_2020, hegre_can_2021}.

Exacerbating the problem, the data used to capture both conflict and the various potential explanatory factors suffers from multiple shortcomings and biases. For many socio-economic and political risk factors of conflict, data varies widely in quality \citep{cederman_predicting_2017, chadefaux_conflict_2017, murphy_promise_2024}, e.g. due to difficulties and expenses connected to measurement, or deliberate misrepresentation by governments. For many countries, subnational and subyearly data is also either not available or lags too far behind reality to be useful for forecasting tasks at the \emph{pgm}-level. Additionally, data on structural risk factors often shows too little variance for meaningful predictions regarding the timing of future conflict \citep{mueller_reading_2018, chadefaux_endogenous_2025}, while other data does not exist at all or only in limited contexts. Consequently, the only fine-grained, high-variance features available often are conflict history features, which tend to have an outsized impact in terms of predictive power \citep{hegre_can_2021, hegre_forecasting_2022-1, mueller_hard_2022, chadefaux_endogenous_2025}. On the one hand, this corresponds to the well-documented issue of conflict recurrence \citep{collier_breaking_2003}, on the other hand conflict likely also proxies many unobserved factors that have already resulted in violence. Current forecasting systems therefore do much better at discovering locations at risk than predicting when conflict is most likely to erupt \citep{mueller_reading_2018, bazzi_promise_2022}.

However, the conflict event datasets that forecasts rely on, not only as the most important feature set but also for the prediction target, suffer from several biases and shortcomings themselves. Mostly based on news reporting, both the accuracy of the information contained and the selection of events covered are potentially affected \citep{earl_use_2004, althaus_total_2022}. While not perfect, information on the events ultimately recorded has been evaluated as reasonably accurate, with errors regarding exact location and number of fatalities within acceptable boundaries for the \emph{pgm} prediction problem \citep{weidmann_accuracy_2015}. In contrast, it is often hard to determine what portion of violence is reported at all and to what extent it is biased exactly, given the difficulty of establishing an accurate ground truth in a conflict context \citep{price_selection_2015, althaus_total_2022}. Selection biases generally depend on the widely varying judgments of newsworthiness and (partisan) preferences by a particular source \citep{davenport_views_2002, baum_filtering_2015, dietrich_known_2020}, while also impacted by logistical factors such as the possibility for information to reach reporters \citep{weidmann_closer_2016, croicu_communication_2017}. In sum, this likely results in considerable variance in data quality across different contexts.

These limitations add up to significant hidden uncertainty in point predictions, with forecasts for a given month and location almost guaranteed to be wrong to some degree.

\subsection{Uncertainty Quantification}
\label{sec:theoryuncertainty}

While UQ has received only limited attention in conflict prediction, a large body of research discusses UQ in machine learning settings. Uncertainty can be conceptually split into aleatoric uncertainty and epistemic uncertainty. Aleatoric uncertainty is caused by inherently stochastic processes and therefore irreducible, while epistemic uncertainty is the result of uncertainties related to the modeling process and can theoretically be minimized by additional data and better (fitted) models -- although this may not always be possible in practice \citep{hullermeier_aleatoric_2021, gruber_sources_2025}. Methods to quantify this uncertainty can be loosely summarized as methods to estimate the predictive distribution, resulting either in probabilistic classification or prediction intervals \citep{cheng_machine_2023, gawlikowski_survey_2023, tyralis_review_2024, gruber_sources_2025}, with  many existing approaches rooted in Bayesian analysis \citep{berger_statistical_1985, lampinen_bayesian_2001}.

Distinguishing between the types of uncertainty is often part of UQ approaches \citep{abdar_review_2021, hullermeier_aleatoric_2021, gruber_sources_2025}. However, \citet{gruber_sources_2025} argue that data-related uncertainties blur the boundaries between aleatoric and epistemic uncertainty, with missing or biased data -- a key challenge for conflict prediction -- resulting in increased aleatoric uncertainty in practice, despite this being a theoretically solvable problem. Given the data-related challenges discussed above, at least some seemingly stochastic aspects of conflict \citep{chadefaux_conflict_2017} may simply be the result of missing data, e.g. on the motivations of key actors, rather than an inherently stochastic process. Hence, focusing on the predictive distribution for UQ as proposed by the VIEWS (Violence \& Impacts Early-Warning System) team \citep{hegre_202324_2025} without trying to disentangle these types of uncertainty constitutes a sufficient first step in line with common practices for UQ in ML \citep{tyralis_review_2024}.

With these considerations in mind, we design a ML-based approach to conflict prediction incorporating UQ. In principle, ML is well-suited to capture the complexity of conflict through its ability to account for non-linear effects and interaction effects between multiple variables \citep{cederman_predicting_2017}. We use tree-based models, which often outperform the more popular deep learning approaches on tabular data \citep{grinsztajn_why_2022}. While deep learning approaches are in principle more capable in big data applications, their potential has so far not materialized in the field of conflict forecasts, likely connected to the relative rarity of conflict occurrence. Our UQ approach specifically addresses two main data-related sources of uncertainty:

First, we exclusively use ML algorithms with the ability to natively output uncertainty estimates. Importantly, given the potential of similar situations leading to widely different outcomes locally and over time, we focus on algorithms with the ability to predict distributions for each \emph{pgm} individually rather than estimating uncertainty globally. 

Second, we implement a hurdle approach to distinguish between zero and non-zero observations. This assumes differences in distributions on either side of a threshold, which is arguably the case in conflict data due to a combination of biases in the data-generation process as well as differences between the determinants of the occurrence of conflict and the determinants of the intensity of conflict \citep{lacina_explaining_2006, hegre_forecasting_2022}. The two-stage approach focusing first on the likely location of violence before predicting its intensity is also well suited to capture the spatial clustering and diffusion patterns of armed conflict \citep{buhaug_contagion_2008, schutte_diffusion_2011, racek_capturing_2025}. 

Additionally, to work around some of the data availability challenges, we explore the selective combination of ``local'' and ``global'' models and its impact on model performance. While local predictions have been found to suffer from some of the same issues as of conflict forecasting as a whole \citep{bazzi_promise_2022}, their integration into larger spatial ensembles is yet untested. Our contribution thus paves the way for the selective inclusion of locally available data sources into larger forecasting systems.

\section{Methods}
\label{sec:methods}

As the models were submitted to the VIEWS prediction challenge \citep{hegre_202324_2025}, we generated predictions for each month in six yearly test windows (2018-2023) and (at the time) true future predictions for July 2024-June 2025.\footnote{The true future predictions are tracked and evaluated at \url{https://viewsforecasting.org/leaderboard/} \citep{views_leaderboard_nodate}.}

\subsection{Data and Modeling Setup}
\label{sec:methodsdata}

With our main focus on modeling strategies, we rely on the data provided by the VIEWS team in the context of the prediction challenge and use the full feature set\footnote{These include information on conflict history, geography, natural resources, population size, climate and vegetation, and to some extent the economy. A codebook is available with the data provided in the context of the VIEWS challenge \citep{hegre_202324_2023}.} in our models. The data covers all PRIO-GRID \citep{tollefsen_prio-grid_2012} cells in Africa and the Middle East, a subset with $N=13,110$ cells of the global $0.5^\circ \times 0.5^\circ$ grid, and are available monthly starting in 1990. As outlined in the prediction challenge call, the target is the number of fatalities from state-based armed conflict events \citep{hegre_202324_2023}, as recorded by the Uppsala Conflict Data Program (UCDP) \citep{sundberg_introducing_2013, davies_organized_2025}. The target is highly zero-inflated, with less than 0.4\% non-zero values in the average month in our training data.

To generate predictions for the whole next year from the available training data, we chose to train separate models for each of the timesteps to predict ($t+3, ..., t+14$) and combine the resulting outputs to a full year of predictions. The data cutoff for each test window is three months before the start, i.e. October 2017 for 2018 predictions.\footnote{This simulates the availability gap present in the data for the true future predictions, where training data was available until April 2024 to predict for the timeframe July 2024 to June 2025. Therefore, predictions need to be generated for the timesteps $t+3, ..., t+14$.} The training data is further limited by the time period we want to predict into the future with this approach. For example, the model generating the predictions for December 2018 ($t+14$) based on data up to October 2017 ($t$) can only be trained on data up to August 2016 ($t-14$), as this is the last month where there is sufficient future information to label the target.

We designed a modular, model-agnostic modeling pipeline to perform the tuning, training, and predicting automatically for the given prediction problem. We include multiple machine learning algorithms, selecting the algorithm which achieves the best performance during cross-validation for each timestep individually. This allows us to integrate and test different machine learning algorithms with minimal effort and means our approach can be easily reused for different prediction problems, with the code available as part of our replication material.

We perform hyperparameter tuning for each timestep once based on the data up to October 2017 ($N=4,378,740$), before the first test window. Our tuning procedure is based on time series cross-validation\citep{petropoulos_forecasting_2022} with a 5-year sliding window through the training data. We employ a Bayesian search approach with the Tree-structured Parzen Estimator (TPE) algorithm using \emph{hyperopt} \citep{bergstra_algorithms_2011, bergstra_making_2013}, to efficiently estimate the best hyperparameters for each prediction timestep model. The cross-validation and scoring within the tuning procedure is implemented in the \emph{scikit-learn} framework \citep{pedregosa_scikit-learn_2011}. After tuning, we fit the model on all available training data for a given test window before generating our predictions from the last available observations three months before the start of the prediction window.

Our tuning metric depends on the modeling step in the hurdle model, with further details provided below. For each chosen base metric, our tuning metric is defined  as the mean performance across the test folds during cross-validation. As we observed a strong tendency to overfit for some parameter combinations, we combine this with a penalty term on deviations between mean train and test fold performance, in order to favor generalizability of the models.\footnote{The formula we use is $M_{tune} = \overline{M}_{test} - 0.5|\overline{M}_{train} - \overline{M}_{test}|$, where higher scores for all $M$ are better, and $\overline{M}$ represents the mean performance across all train and test folds in the cross-validation, respectively. The tuning metric $M_{tune}$ is then maximized by the tuning algorithm.}

\subsection{Hurdle Approach}
\label{sec:methodshurdle}

Our overarching modeling approach is a variation on the principle of hurdle models. Hurdle models \citep{mullahy_specification_1986} are a combination of two modeling steps: a first step consisting of a classifier determining whether the hurdle is reached, trained on all available training data, and a second step consisting of a regressor determining the predicted value, trained on only the subsection of the training data where the target is above the hurdle. We perform tuning and prediction for each of these two steps separately to allow for custom combinations of global and local predictions (see \ref{sec:methodshurdleensemble}).

\subsubsection{Classifier}
\label{sec:methodsclassifier}

Our classifiers are trained on a binary variable indicating whether or not any fatalities occurred in a given \emph{pgm}. We include two different probabilistic classifiers in our modeling pipeline: Random Forests \citep{breiman_random_2001} and eXtreme Gradient Boosting \citep{chen_xgboost_2016} models.\footnote{We also tested logit models as a “simple” alternative approach, which performed worse by a factor of 2-3 on average.}

\begin{itemize}
    \item \textbf{Random Forests (RF)} aggregate the results of a multitude of decision trees. Each tree in the forest is built from a different sample of data, using a technique called bootstrap aggregating, or bagging. Additionally, Random Forests employ random feature selection, where each split in a tree considers only a random subset of features. 
    \item \textbf{eXtreme Gradient Boosting (XGB)} employs a regularized learning objective that balances model complexity and predictive accuracy. The system utilizes gradient tree boosting, where the model is trained in an additive manner, incrementally improving the predictions by minimizing a loss function using second-order gradient statistics. Models are built sequentially, correcting the errors of previous models.
\end{itemize}

Our tuning metric of choice for classification is the area under the precision-recall-curve. This metric is well suited for zero-inflated classification tasks, as it does not take into account whether zeroes are predicted correctly -- which we argue is of little interest given the relative scarcity of violence \citep[also see][]{saito_precision-recall_2015}. Tuning performance for both model types is fairly similar, with RF performing slightly better in all global models, while XGB is favored for most local models.

\subsubsection{Regressor}
\label{sec:methodsregressor}

To estimate uncertainty around predictions, we rely on regressors designed to output distributions directly rather than trying to estimate distributions around point predictions ex post. We include three different tree-based distributional regressors in our modeling pipeline: Quantile Regression Forests \citep{meinshausen_quantile_2006}, Distributional Random Forests \citep{cevid_distributional_2022} and Natural Gradient Boosting for probabilistic regression \citep{duan_ngboost_2019}.

\begin{itemize}
    \item \textbf{Quantile Regression Forests (QRF)} extend the RF methodology to estimate conditional quantiles. Like RF, the QRF algorithm involves growing an ensemble of decision trees using a randomized node and split point selection process. Unlike traditional RFs, which only retain the mean response in each leaf, QRF retains all observed responses, enabling the estimation of the entire conditional distribution. The algorithm calculates the conditional quantile by averaging the weighted distribution of observed responses, with weights derived from the original RF methodology. We use evenly spaced quantile steps to generate the samples for our predictions with this algorithm.
    \item \textbf{Distributional Random Forests (DRF)} are another extension of the traditional RF framework used to estimate the entire conditional distribution of univariate or multivariate responses. The methodology involves constructing trees that split data points based on a novel criterion derived from the Maximum Mean Discrepancy (MMD) statistic, which measures differences in distributions rather than just differences in means. This splitting criterion is applied recursively to ensure that the distributions in the resulting child nodes are as homogeneous as possible. Each tree in the forest is grown to optimize this distributional metric. The final forest model uses a weighted combination of trees to estimate the full conditional distribution of the response variables. This approach allows DRF to adaptively weight training data points based on their relevance to the prediction, providing a robust and flexible method for modeling complex dependencies.
    \item \textbf{Natural Gradient Boosting (NGB)} for probabilistic regression extends gradient boosting to the estimation of probability distributions. This involves boosting the parameters of a specified parametric distribution using a natural gradient, which corrects the training dynamics for more stable and efficient learning. The algorithm integrates three modular components: a simple base learner, a parametric probability distribution, and a proper scoring rule. The natural gradient is employed to optimize the parameters of the conditional distribution, ensuring that the updates are invariant to reparameterization and efficiently exploit the curvature of the score in distributional space. We use decision trees as the base learner, log-normal probability distributions and the Continuous Ranked Probability Scores (CRPS) as the scoring rule.
\end{itemize}

Following the principle of hurdle models, we train our regression models only on \emph{pgms} with non-zero targets while still generating predictions for all \emph{pgms}. As our tuning metric, we use the challenge’s main metric, the CRPS \citep[see equation \ref{eq:1},][]{hegre_202324_2025}. In line with the maximum sample size allowed by the prediction challenge, we set our regression models to output a forecast sample with $N_{sample}=1,000$, drawn from the predicted distribution. This results in a wider range of possible values, ensuring the inclusion of low-probability outcomes. NGB performed best during tuning, being chosen in 75\% of global models and 80\% of local models, with the other two regressors only chosen occasionally.

\begin{landscape}
    \begin{figure}
    	\centering
        \includegraphics[width=1\linewidth]{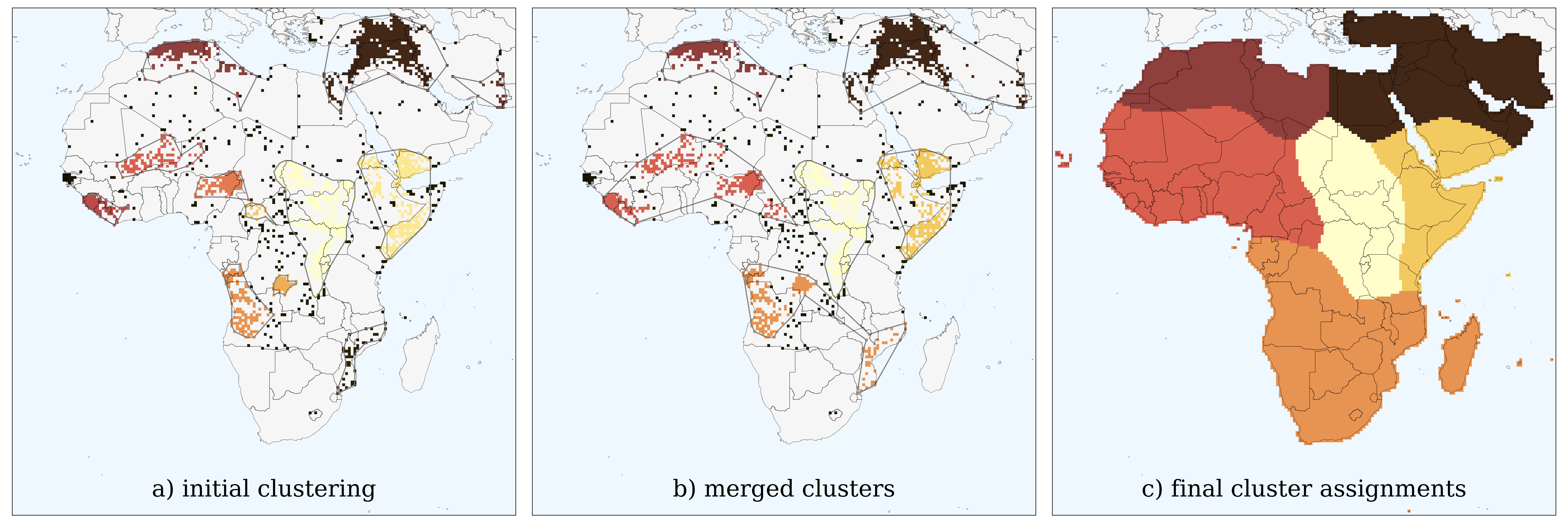}
    	\caption{Visualization of the process to create the geographic regions for local models at three steps: a) clusters created by the manually tuned HDBSCAN algorithm with corresponding polygons; b) updated clusters after combining smaller clusters with corresponding polygons; c) final contiguous regions with grid-cells without violence integrated. Grid cells shown in a and b are those experiencing any violence in the training data (1990-2017). Black grid cells displayed in a and b are not assigned to any clusters by HDBSCAN initially.}
    	\label{fig:fig1}
    \end{figure}
\end{landscape}

\subsubsection{Quasi-Hurdle Ensemble}
\label{sec:methodshurdleensemble}
Hurdle model point predictions are usually generated via a simple multiplication of the output of the classifier and the predicted value of the regressor. Given the samples drawn from the predictive distribution, a multiplication of the classification probability with each draw in the samples would result in non-integer predictions, which is not in line with the nature of fatality counts. At the same time, our tree-based regressors trained only on non-zero targets never produce zero predictions\footnote{The only exception here is NGB which does also predict zeroes. For consistency with the other algorithms we replace all zeros with one as the lowest possible non-zero value in the NGB predictions.}, a multiplication would therefore likely overestimate the probability of violence occurring. While both issues could be partially addressed with rounding, we opt to instead treat the classification probability as the percentage of the ensemble sample taken from the non-zero predictions via a random draw, with the remaining share of the $1,000$ draws filled with zero values. In testing, this also performed better than the multiplicative approach.\footnote{We also tested a more "traditional" hurdle combination, selecting either all-zero samples or the full non-zero sample based on the classification probability and a threshold, which performed significantly worse.}

\subsection{Local Models}
\label{sec:methodslocal}

In addition to the ``global'' approach described above, we explore the integration of multiple ``local'' models encompassing smaller geographic regions. This has two potential benefits: first, it reduces the spatial coverage requirements and thus increases the pool of available datasets for a prediction problem. Second, it accounts for potential systematic differences across regions, either due to differences in conflict patterns or due to differences in data generation. Focusing on potential differences in conflict patterns, we create custom contiguous geographic regions based on the spatial distribution of grid cells with any recorded fatalities in the training data (1990-2017) using the HDBSCAN clustering algorithm \citep{campello_density-based_2013}.\footnote{We tested two additional versions of generating clusters: One with clusters created via an alternative clustering algorithm, DBSCAN, and a similar manual tuning of clustering parameters, and one with clusters corresponding to the United Nations Statistics Division sub-regions more aligned to the data coverage problem, with grid cells assigned to countries based on a majority rule. Both performed only slightly worse in testing. While we did not include them in our final model run and evaluation, corresponding models can still be produced with our replication code.} We manually tested parameter combinations for the clustering algorithm until inspections of the results yielded groupings, which plausibly corresponded to visually discernible patterns, resulting in eleven clusters. To ensure sufficient non-zero training data in each cluster for the hurdle regression models, we further reduce this down to six clusters by iterating over the clusters and combining smaller clusters with their nearest neighbors based on centroid distance of polygons drawn around each cluster, requiring a minimum of $1,000$ \emph{pgms} with non-zero fatalities.

To subsequently create contiguous regions including any cells not experiencing conflict, we first draw new polygons around the combined grid cells of each cluster. Cells remaining outside these polygons were assigned to the cluster with the nearest boundary to the cell center. The procedure and resulting clusters are visualized in Figure \ref{fig:fig1}. We subsequently train separate local models for each of the clusters, following the same procedure as with the global models described above. Each grid cell is therefore assigned not only to exactly one cluster but also to one corresponding local set of models. Combining the predictions from all local models yields predictions for the whole geographical area of interest.

\subsubsection{Global-Local Ensemble}
\label{sec:methodsgloballocalensemble}
We create two spatial ensembles in addition to the original global hurdle model: a local-only hurdle model following the same approach, simply concatenating the predictions from all local models along their spatial IDs, and a global-local ensemble. The latter is constructed by selectively picking and choosing both classification and regression components from either the global or the local model, creating an ensemble of global and local model outputs. We select the components by comparing the combined performance of the hurdle ensemble for each of the four possible combinations of classification and regression predictions on a cluster-by-cluster basis, selecting the global-local combination for each cluster which performed best across the three years prior to a given prediction window.\footnote{Repeating this with only one or two years of prior data resulted in worse performance of the combined prediction. This is likely connected to a fairly high volatility in performance across years, with prior performance not correlated enough to future performance. An increase to five years of prior data as the selection basis did not lead to meaningful improvements.} Finally, the resulting predictions for each cluster are spatially concatenated analogous to the local-only model.

\begin{figure}
    \centering
    \includegraphics[width=1\linewidth]{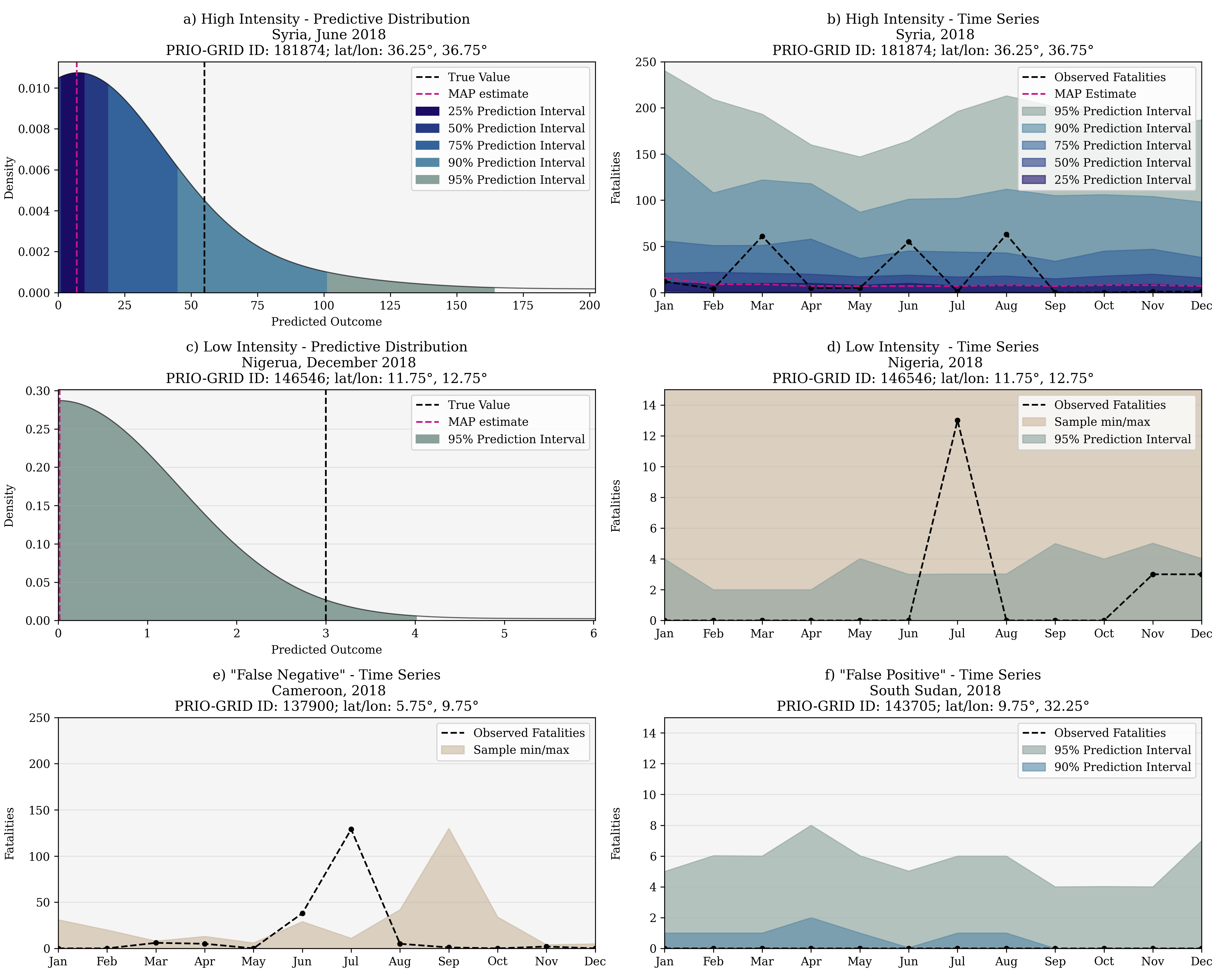}
    \caption{Predictive distributions and their development over time for example grid cells in the global model's 2018 predictions, compared to the observed number of fatalities and a maximum a posteriori (MAP) estimate for the distributions. Figures a and c show the distributions for single \emph{pgms}, estimated via Gaussian kernel density estimation from the forecast samples, with X-axes limited to below the 98\% quantile. Figures b, d-f show the times series for the whole test window, visualized through selected quantile-based prediction intervals. MAP estimates are calculated as the maximum of the estimated probability density function.}
    \label{fig:fig2}
\end{figure}

\section{Results and Evaluation}
\label{sec:results}

To illustrate what kind of distributions our models\footnote{For simplicity, we subsequently use the term ``model'' in this context to refer to the respective ensemble specification used to generate our global-only, local-only, or global-local predictions.} predict, Figure \ref{fig:fig2} compares observed fatalities during the first test window with our global model predictions for four example grid cells representing typical patterns: A high-intensity case (a, b) and a low-intensity case (c, d) we capture fairly well, and two “misses”: a “false negative” case (e) where our predictions showed only a minuscule chance for violence but in fact a significant number of fatalities occurred, and a “false positive” case (f) where our predictions saw a small but noticeable chance for violence usually seen in low intensity cases, but no violence occurred.

As becomes immediately apparent, our predictions are heavily right-skewed with long tails and a large number of zeroes (a, c), indicating mostly low probabilities in the classification step. Notably, this zero-inflatedness mirrors the distribution of observed fatalities across \emph{pgms}. The predictions capture general conflict intensity well in most cases, but our hurdle approach only leads to maximum a posteriori (MAP) estimates (see \ref{sec:resultsperformance}) above zero for the highest-intensity predictions (a, b). Instead, higher conflict intensity is represented through a greater likelihood of non-zero fatalities. In line with other prediction systems \citep{mueller_reading_2018, bazzi_promise_2022}, our models are better at predicting the locations of fatalities than their timing, with variance in the predictive distributions over time rarely matching observed fatalities (b, d). However, with the distributional approach, the spikes of violence observed in real-world data can be treated simply as lower likelihood events, but within the range of possible values. 

Typical misses include “false negative” cases, where significant violence occurred despite our sample containing more than 99\% zeroes, such as in the Cameroon example, located close to the Nigerian border (e). However, our models partially account for this by always assuming a small base probability of violence with the samples never being all-zero, which is the result of our hurdle combination approach and a non-zero minimum probability from the classification step. In addition, we see a number of “false positives” (f), where a sizable share of our samples is non-zero but this predicted risk does not materialize. These predictions generally are part of a spatial cluster of higher-likelihood predictions with fatalities observed only in some of its grid cells. The model thus assumes elevated risk in cells near active conflict and predictions are similar to those where low-intensity violence did occur (d), without being able to reliably identify exact locations.

\subsection{Aggregate Performance}
\label{sec:resultsperformance}

\begin{table}
\centering
{
    \setlength{\dashlinedash}{2pt}
    \setlength{\dashlinegap}{2pt}
    \setlength{\extrarowheight}{2pt}
    \renewcommand{\ttdefault}{pcr}
    \small\ttfamily
    
    \begin{tabular}{|l|l!{\vrule width 1.2pt}r|r|r:r|r|}
    \hline
    \rowcolor{gray!30} \textbf{Model} & \textbf{Year} & \textbf{CRPS} & \textbf{IGN} & \textbf{MIS} & \textbf{ MSE} & \textbf{ MAE} \\
    \boldhline
    \rowcolor{cdark1} Global & 2018 & \textbf{0.1300} & \textbf{0.0681} & \textbf{2.1646} & \cellcolor{cdark2} 62.396 & \cellcolor{cdark2} 0.1487 \\
    \rowcolor{cdark1} Local & 2018 & 0.1376 & 0.0824 & 2.4546 & \cellcolor{cdark2} \textbf{59.707} & \cellcolor{cdark2} 0.1516 \\
    \rowcolor{cdark1} Global-local & 2018 & 0.1324 & 0.0691 & 2.2266 & \cellcolor{cdark2} 61.682 & \cellcolor{cdark2} 0.1538 \\
    \rowcolor{cdark1} All-zero & 2018 & 0.1444 & 0.0916 & 2.8883 & \cellcolor{cdark2} 65.815 & \cellcolor{cdark2} \textbf{0.1444} \\
    \rowcolor{cdark1} Poisson (last) & 2018 & 0.3860 & 0.1177 & 7.1488 & \cellcolor{cdark2} 169.45 & \cellcolor{cdark2} 0.4020 \\
    \rowcolor{cdark1} Conflictology & 2018 & 0.1919 & 0.8589 & 2.8345 & \cellcolor{cdark2} 86.968 & \cellcolor{cdark2} 0.2232 \\
    \rowcolor{cdark1} Conf. neighbors & 2018 & 0.1473 & 0.1770 & 3.0622 & \cellcolor{cdark2} 64.951 & \cellcolor{cdark2} 0.1583 \\
    \rowcolor{cdark1} Bootstrap 240 & 2018 & 0.1443 & 0.0925 & 2.8883 & \cellcolor{cdark2} 65.815 & \cellcolor{cdark2} \textbf{0.1444} \\
    \boldhline
    
    \rowcolor{clight1} Global & 2019 & \textbf{0.1010} & \textbf{0.0664} & \textbf{1.5599} & \cellcolor{clight2} 16.665 & \cellcolor{clight2} \textbf{0.1153} \\
    \rowcolor{clight1} Local & 2019 & 0.1040 & 0.0805 & 1.8239 & \cellcolor{clight2} \textbf{16.515} & \cellcolor{clight2} 0.1156 \\
    \rowcolor{clight1} Global-local & 2019 & 0.1011 & 0.0668 & 1.6023 & \cellcolor{clight2} 16.603 & \cellcolor{clight2} 0.1173 \\
    \rowcolor{clight1} All-zero & 2019 & 0.1154 & 0.0944 & 2.3089 & \cellcolor{clight2} 17.239 & \cellcolor{clight2} 0.1154 \\
    \rowcolor{clight1} Poisson (last) & 2019 & 0.1442 & 0.1050 & 2.6166 & \cellcolor{clight2} 18.224 & \cellcolor{clight2} 0.1515 \\
    \rowcolor{clight1} Conflictology & 2019 & 0.1184 & 0.8561 & 1.8887 & \cellcolor{clight2} 17.444 & \cellcolor{clight2} 0.1275 \\
    \rowcolor{clight1} Conf. neighbors & 2019 & 0.1068 & 0.1755 & 1.8786 & \cellcolor{clight2} 16.717 & \cellcolor{clight2} 0.1204 \\
    \rowcolor{clight1} Bootstrap 240 & 2019 & 0.1154 & 0.0951 & 2.3089 & \cellcolor{clight2} 17.239 & \cellcolor{clight2} 0.1155 \\
    \boldhline
    
    \rowcolor{cdark1} Global & 2020 & \textbf{0.1180} & 0.0743 & 1.9011 & \cellcolor{cdark2} 16.812 & \cellcolor{cdark2} \textbf{0.1311} \\
    \rowcolor{cdark1} Local & 2020 & 0.1221 & 0.0872 & 2.2054 & \cellcolor{cdark2} 16.708 & \cellcolor{cdark2} 0.1338 \\
    \rowcolor{cdark1} Global-local & 2020 & 0.1181 & \textbf{0.0742} & \textbf{1.9001} & \cellcolor{cdark2} 16.796 & \cellcolor{cdark2} 0.1315 \\
    \rowcolor{cdark1} All-zero & 2020 & 0.1319 & 0.1077 & 2.6374 & \cellcolor{cdark2} 17.218 & \cellcolor{cdark2} 0.1319 \\
    \rowcolor{cdark1} Poisson (last) & 2020 & 0.1646 & 0.1163 & 2.9928 & \cellcolor{cdark2} 18.747 & \cellcolor{cdark2} 0.1725 \\
    \rowcolor{cdark1} Conflictology & 2020 & 0.1275 & 0.8599 & 2.0731 & \cellcolor{cdark2} \textbf{16.547} & \cellcolor{cdark2} 0.1365 \\
    \rowcolor{cdark1} Conf. neighbors & 2020 & 0.1230 & 0.1818 & 2.1152 & \cellcolor{cdark2} 16.932 & \cellcolor{cdark2} 0.1334 \\
    \rowcolor{cdark1} Bootstrap 240 & 2020 & 0.1317 & 0.1072 & 2.6374 & \cellcolor{cdark2} 17.218 & \cellcolor{cdark2} 0.1319 \\
    \boldhline
    
    \rowcolor{clight1} Global & 2021 & 0.9246 & \textbf{0.0843} & 17.9788 & \cellcolor{clight2} 81844.5 & \cellcolor{clight2} 0.9383 \\
    \rowcolor{clight1} Local & 2021 & 0.9293 & 0.0964 & 18.3496 & \cellcolor{clight2} 81844.5 & \cellcolor{clight2} 0.9405 \\
    \rowcolor{clight1} Global-local & 2021 & \textbf{0.9243} & \textbf{0.0843} & 17.9785 & \cellcolor{clight2} 81844.2 & \cellcolor{clight2} \textbf{0.9377} \\
    \rowcolor{clight1} All-zero & 2021 & 0.9398 & 0.1188 & 18.7961 & \cellcolor{clight2} 81844.9 & \cellcolor{clight2} 0.9398 \\
    \rowcolor{clight1} Poisson (last) & 2021 & 0.9703 & 0.1286 & 19.0801 & \cellcolor{clight2} 81843.2 & \cellcolor{clight2} 0.9793 \\
    \rowcolor{clight1} Conflictology & 2021 & 0.9302 & 0.8648 & \textbf{17.8700} & \cellcolor{clight2} \textbf{81843.0} & \cellcolor{clight2} 0.9426 \\
    \rowcolor{clight1} Conf. neighbors & 2021 & 0.9279 & 0.1893 & 18.1056 & \cellcolor{clight2} 81844.5 & \cellcolor{clight2} 0.9414 \\
    \rowcolor{clight1} Bootstrap 240 & 2021 & 0.9396 & 0.1175 & 18.7961 & \cellcolor{clight2} 81844.9 & \cellcolor{clight2} 0.9398 \\
    \boldhline
    
    \rowcolor{cdark1} Global & 2022 & 1.1274 & 0.0834 & 22.2467 & \cellcolor{cdark2} 98555.7 & \cellcolor{cdark2} 1.1396 \\
    \rowcolor{cdark1} Local & 2022 & 1.1289 & 0.0951 & 22.4550 & \cellcolor{cdark2} 98555.4 & \cellcolor{cdark2} 1.1402 \\
    \rowcolor{cdark1} Global-local & 2022 & \textbf{1.1263} & \textbf{0.0832} & 22.2353 & \cellcolor{cdark2} 98553.7 & \cellcolor{cdark2} 1.1403 \\
    \rowcolor{cdark1} All-zero & 2022 & 1.1375 & 0.1199 & 22.7494 & \cellcolor{cdark2} 98560.0 & \cellcolor{cdark2} \textbf{1.1375} \\
    \rowcolor{cdark1} Poisson (last) & 2022 & 1.4565 & 0.1453 & 28.5266 & \cellcolor{cdark2} 98770.9 & \cellcolor{cdark2} 1.4734 \\
    \rowcolor{cdark1} Conflictology & 2022 & 1.1419 & 0.8669 & \textbf{22.2770} & \cellcolor{cdark2} \textbf{98504.3} & \cellcolor{cdark2} 1.1532 \\
    \rowcolor{cdark1} Conf. neighbors & 2022 & 1.1311 & 0.1896 & 22.4754 & \cellcolor{cdark2} 98547.9 & \cellcolor{cdark2} 1.1442 \\
    \rowcolor{cdark1} Bootstrap 240 & 2022 & 1.1373 & 0.1181 & 22.7494 & \cellcolor{cdark2} 98560.0 & \cellcolor{cdark2} \textbf{1.1375} \\
    \boldhline
    
    \rowcolor{clight1} Global & 2023 & \textbf{0.2147} & \textbf{0.0863} & \textbf{3.9807} & \cellcolor{clight2} 163.26 & \cellcolor{clight2} 0.2275 \\
    \rowcolor{clight1} Local & 2023 & 0.2207 & 0.0988 & 4.2879 & \cellcolor{clight2} \textbf{163.17} & \cellcolor{clight2} 0.2565 \\
    \rowcolor{clight1} Global-local & 2023 & 0.2175 & 0.0867 & 4.1036 & \cellcolor{clight2} 163.52 & \cellcolor{clight2} 0.2444 \\
    \rowcolor{clight1} All-zero & 2023 & 0.2236 & 0.1210 & 4.4723 & \cellcolor{clight2} 163.33 & \cellcolor{clight2} \textbf{0.2236} \\
    \rowcolor{clight1} Poisson (last) & 2023 & 9.7500 & 0.1507 & 193.97 & \cellcolor{clight2} 1134237 & \cellcolor{clight2} 9.7807 \\
    \rowcolor{clight1} Conflictology & 2023 & 0.5237 & 0.8686 & 13.218 & \cellcolor{clight2} 369.37 & \cellcolor{clight2} 0.3576 \\
    \rowcolor{clight1} Conf. neighbors & 2023 & 0.2499 & 0.1922 & 4.0334 & \cellcolor{clight2} 163.39 & \cellcolor{clight2} 0.2338 \\
    \rowcolor{clight1} Bootstrap 240 & 2023 & 0.2234 & 0.1196 & 4.4723 & \cellcolor{clight2} 163.33 & \cellcolor{clight2} \textbf{0.2236} \\
    \hline
    
    \end{tabular}
}
\vspace{1em}
\caption{Average model and benchmark metrics across all six yearly test windows. Best results for each year are marked in bold. Lower scores signify better performance. Note that the CRPS is always equal to the MAE in the case of all-zero predictions.}
\label{tab:table1}
\end{table}

To evaluate the overall performance of our three models, we compare them to several benchmarks provided by the VIEWS team across the six yearly test windows. Those are two naive benchmarks and three “conflictology” benchmarks based on medium- to long-term conflict history. The naive benchmarks are samples drawn from a Poisson distribution centered around the last observed values for each grid cell and predictions with only zero values. The “conflictology” benchmarks all treat historic fatality counts as draws from the predictive distribution to generate forecasts. The first benchmark (“conflictology”) uses fatality counts from a specific grid cell during the previous 12 months for the respective prediction window (12 draws). The second benchmark (“conflictology neighbors”) follows the same principle, but uses the combined conflict history of the grid cell and its immediate neighbors (108 draws). The third benchmark (“bootstrap 240”) combines $1,000$ random draws from the grid cell’s conflict history of the last 240 months. All also adhere to the two-month gap between training and test data. For instance, the “conflictology” samples for all months in 2018 are based on the observed fatalities from November 2016 to October 2017 \citep{hegre_202324_2025}.\footnote{Information in addition to \citet{hegre_202324_2025} based on the source code at \url{https://github.com/prio-data/prediction_competition_2023/blob/a45796ce8d1ffdd82e879e05c46d90c58b460a66/benchmark.py}.}

We base our evaluation on the challenge’s main metric, the CRPS, and the two additional metrics specified in the prediction challenge: an adjusted, binned version of the Ignorance Score (IGN) (also called Log Score) and the Mean Interval Score (MIS).\footnote{For consistency with \citet{hegre_202324_2025}, we follow the convention of referring to the \emph{Mean} Interval Score rather than simply the Interval Score in the text, denoting that it is averaged across multiple \emph{pgms} for the evaluation, while other scores are referred to without this prefix despite using the same averaging procedure (see equation \ref{eq:4}} All scores evaluate samples from a distribution, with lower scores representing better predictions and perfect predictions corresponding to a score of zero for CRPS and MIS, and 0.014 for IGN (see equations \ref{eq:5} - \ref{eq:7}). 

For an observation $y_i$ and its corresponding predictive distribution $F_i(.)$, the CRPS is defined as
\begin{equation}
    CRPS(F_i, y_i) = \int_{-\infty}^{\infty} \left( F_i(x) - \mathbbm{1}(x-y_i) \right)^2 \, dx, 
    \quad \text{where} \quad 
    \mathbbm{1}(z) = \begin{cases}
        1 & \text{if } z \geq 0 \\ 
        0 & \text{otherwise} 
    \end{cases}\text{.}
    \label{eq:1}
\end{equation}

Following VIEWS, the adjusted, binned IGN is defined as 
\begin{equation}
    ab\text{-}IGN(F_i, y_i) = -\log_2 \left(\frac{ n_{b_i} + \omega }{ N_{sample} + N_{bins}\,\omega }\right),
    \label{eq:2}
\end{equation} where $n_{b_i}$ denotes the count of draws in bin $b_i \in \mathcal{B}$ with $y_i \in b_i$, $N_{sample}$ is the total number of draws from $F_i$,\linebreak $N_{bins}$ is the number of individual bins, and $\omega$ is a constant adjustment value. In line with the challenge, we use $N_{sample} = 1,000$, $\omega = 1$, and the bins $\mathcal{B}=\{\{0\},[1,2],[3,5],[6,10],[11,25],[26,50],[51,100],$ $[101,250],[251,500],[501,1000],[1001,\infty)\}$, i.e. $N_{bins} = 11$.

The Interval Score (IS) is defined as 
\begin{equation}
    IS(F_i, y_i) = (U_i - L_i) + \frac{2}{q}(L_i - y_i)\times\mathbbm{1}(L_i > y_i) + \frac{2}{q}(y_i - U_i)\times\mathbbm{1}(y_i > U_i),
    \label{eq:3}
\end{equation} with $U_i$ denoting the upper $1-\frac{q}{2}$ quantile, $L_i$ denoting the lower $\frac{q}{2}$ quantile of $F_i$, and the same indicator function $\mathbbm{1}$ as in (\ref{eq:1}). The challenge uses a 90\% prediction interval, which means $q = 0.1$.\footnote{CRPS is calculated using the \emph{xskillscore} Python package \citep{bell_xskillscore_2021}. See \url{https://github.com/prio-data/prediction_competition_2023} for the challenge-specific implementations of the MIS and IGN scores.} 

Yearly scores for all metrics are calculated as the mean of the score over all incuded \emph{pgms} \citep{hegre_202324_2025}, for example
\begin{equation}
    \overline{CRPS} = \frac{1}{N} \sum_{i=1}^{N} CRPS(F_i, y_i).
    \label{eq:4}
\end{equation}

For potential comparisons with point prediction models, we also report Mean Squared Error (MSE) and Mean Absolute Error (MAE) values based on MAP estimates. We use Gaussian kernel density estimation (KDE) to estimate the probability density function (PDF), from which we select the maximum.

The evaluation results for our models and the benchmarks are reported in Table 1. Across all test windows, our three models are reliably able to beat the benchmarks, with very few exceptions. Our global and global-local models perform about the same, with the latter even pulling ahead in two to three out of the six test windows, depending on the metric. The local model is only slightly worse. While this means we are unable to exploit systematic differences across conflict contexts to improve predictive performance, this still opens avenues for the selective inclusion of data with limited geographic availability. The strong performance of our models is confirmed by the VIEWS team's evaluation of the true future forecasts, in which our models ranked first among all contributions at the \emph{pgm} level \citep{views_leaderboard_nodate}.

\subsection{Uncertainty in Model Evaluation}
\label{sec:resultsuncertainty}

\begin{table}[H]
\centering
{
    \setlength{\dashlinedash}{2pt}
    \setlength{\dashlinegap}{2pt}
    \setlength{\extrarowheight}{2pt}
    \renewcommand{\ttdefault}{pcr}
    \small\ttfamily
    
    \begin{tabular}{|l!{\vrule width 1.2pt}r|r!{\vrule width 1.2pt}r|r!{\vrule width 1.2pt}r|r|}
    \hline
    \rowcolor{gray!30} 
     & \multicolumn{2}{c!{\vrule width 1.2pt}}{\textbf{CRPS}} & \multicolumn{2}{c!{\vrule width 1.2pt}}{\textbf{MIS}} & \multicolumn{2}{c|}{\textbf{IGN}} \\
    \rowcolor{gray!30} 
     \textbf{Model} & \textbf{Fatalities} & \textbf{\shortstack{Non-zero\\pgms}} & \textbf{Fatalities} & \textbf{\shortstack{Non-zero\\pgms}} & \textbf{Fatalities} & \textbf{\shortstack{Non-zero\\pgms}} \\
    \boldhline
    
    \rowcolor{cdark1} Global & \cellcolor{clight1} 0.99998 & 0.66297 & \cellcolor{clight1} 0.99990 & 0.66577 & \cellcolor{clight1} 0.64328 & 0.98190 \\ 
    \rowcolor{cdark1} All-zero & \cellcolor{clight1} 1.00000 & 0.66025 & \cellcolor{clight1} 1.00000 & 0.66025 & \cellcolor{clight1} 0.66025 & 1.00000 \\ 
    \rowcolor{cdark1} Poisson (last) & \cellcolor{clight1} -0.10729 & 0.54045 & \cellcolor{clight1} -0.10759 & 0.54067 & \cellcolor{clight1} 0.52381 & 0.84532 \\ 
    \rowcolor{cdark1} Conflictology & \cellcolor{clight1} 0.96532 & 0.78414 & \cellcolor{clight1} 0.91448 & 0.84714 & \cellcolor{clight1} 0.60015 & 0.92618 \\
    \rowcolor{cdark1} Conf. neighbors & \cellcolor{clight1} 0.99957 & 0.67415 & \cellcolor{clight1} 0.99951 & 0.64671 & \cellcolor{clight1} 0.62442 & 0.97840 \\ 
    \rowcolor{cdark1} Bootstrap 240 & \cellcolor{clight1} 1.00000 & 0.66016 & \cellcolor{clight1} 1.00000 & 0.66025 & \cellcolor{clight1} 0.65347 & 0.99991 \\ 
    \hline

    \end{tabular}
}
\vspace{1em}
\caption{Correlation between evaluation scores and number of fatalities / number of non-zero \emph{pgms} across the six test windows for the global model and the benchmarks. Note that the perfect correlation for all-zero samples with fatalities for CRPS and MIS, and with non-zero \emph{pgms} for IGN is expected behavior. The perfect correlation of the Bootstrap 240 benchmarks with fatalities for the CRPS is the result of rounding, while for the MIS it is the result of the 90\% prediction interval used \citep{hegre_202324_2025} in combination with the composition of the individual samples. As the samples never contain more than 1.9\% non-zero draws for this benchmark in a given test window, the intervals only contain all-zero predictions.}
\label{tab:table2}
\end{table}

Taking a closer look, differences in performance between both our own models and our models’ improvements over most benchmarks are miniscule, and could conceivably be caused by random components in the modeling process or simply be irrelevant in practice. This warrants a deeper examination of the scoring functions and their properties. First, it should be noted that all three scores in aggregate cannot be compared across the different years, since they are not dependent on the distributions of the underlying data. Furthermore, for the special case of an all-zero forecast, i.e. a forecast drawn from a predictive distribution $F_i^0$ that places all probability mass on zero $P(Y_i=0)=1$, with $F_i^0(0) = 1$, and the highly-common $y_i=0$ observation, individual scores (\ref{eq:1}), (\ref{eq:2}), and (\ref{eq:3}) result in
\begin{equation}
    CRPS(F_i^0, 0) = 0
    \label{eq:5}
\end{equation} for the CRPS,
\begin{equation}
    ab\text{-}IGN(F_i^0, 0) = -\log_2(\frac{1001}{1011})\approx 0.014, \text{ since } n_{b_{i}} = 1000 \text{ with } y_i = 0 
    \label{eq:6}
\end{equation} for the adjusted, binned IGN, and
\begin{equation}
    IS(F_i^0, 0) = 0
    \label{eq:7}
\end{equation} for the MIS, respectively.

The highly zero-inflated nature of our prediction task means only a tiny percentage of predictions actually impact the aggregate scores when the means across all relevant \emph{pgms} are calculated (\ref{eq:4}), as $N$ remains constant. Assuming most zeroes are predicted "correctly" with all-zero forecasts, the low individual scores (\ref{eq:5} - \ref{eq:7}) correspondingly result in very low mean scores, with the informative range for each score becoming very small. Illustrating this effect, mean CRPS and MIS scores are almost perfectly correlated with the number fatalities across the different years, while mean IGN scores are strongly correlated with the number of non-zero \emph{pgms} for most models and benchmarks (Table \ref{tab:table2}).

\begin{figure}
    \centering
    \includegraphics[width=.8\linewidth]{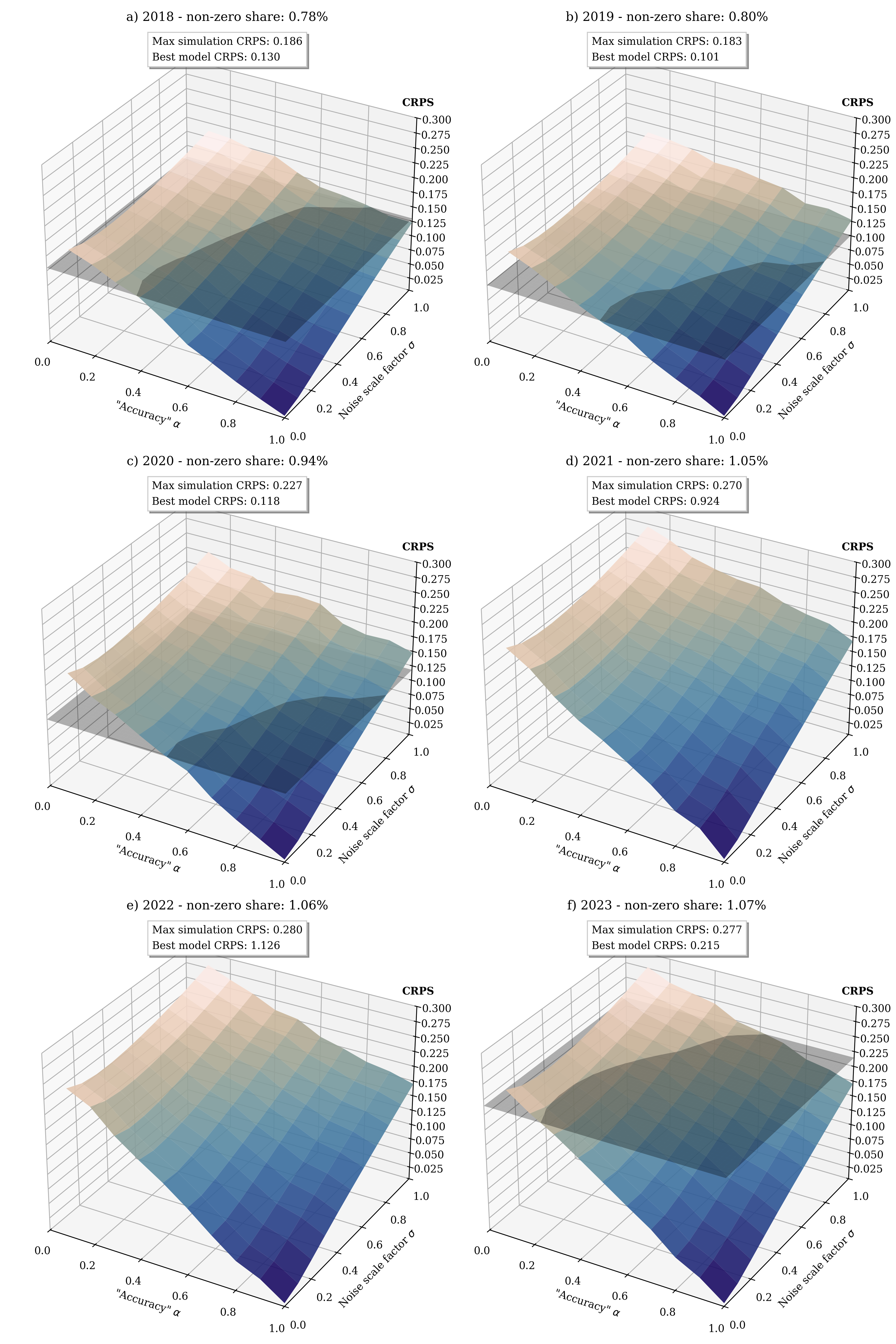}
    \caption[Value ranges of CRPS scores with varying noise and “accuracy” based on simulated data.]{Value ranges of CRPS scores with varying noise and “accuracy” based on simulated data. Simulated observations are created with zero-inflatedness matching the six test windows and values drawn from an estimated PDF based on real non-zero observations $< 1,000$ fatalities. Simulated predictions are drawn from a Poisson distribution with mean and variance equal to the corresponding simulated observation. “Accuracy” ($\alpha$) $< 1$ means a share $1-\alpha$ of non-zero predictions was switched with all-zero predictions. The Noise scale factor ($\sigma$) means observations were shifted by $\varepsilon = 50\sigma u$ before creating the samples, with $-1 < u < 1$ drawn from a uniform distribution. The dark planes represents the scores for our best model in the respective test window.}
    \label{fig:fig3}
\end{figure}

Given this, it is essential to understand the informative ranges of the scores to properly evaluate our models. We use simulated data to experimentally explore the impact of prediction accuracy for the degree of zero-inflation across all test windows. For our simulated observations $z_i$, we create samples matching the size ($N=157{,}320$) and the share $p$ of non-zero \emph{pgms} ($0.0078 < p < 0.0107$, corresponding to $1,277 < N^{+} < 1{,}683$ \emph{pgms}) for each year. We draw the simulated non-zero values $z_i^{+}$ based on a PDF estimated via Gaussian KDE from all observed values $0 < y_i < 1,000$ in the training data, thus controlling for extreme outliers. We multiply all negative values in this sample with -1 to ensure valid fatality numbers. Next, simulated predictions are created based on these simulated observations. “Perfect” predictive distributions are created for each simulated observation $z_i$ by drawing samples with size $N_{sample}=1,000$ from a Poisson distribution, setting the expectation $\lambda$ to the simulated actual value, i.e. $\lambda_i = z_i$ \citep[see][]{hegre_202324_2023}. Finally, we introduce errors to these predictions in two ways: First, we vary the “accuracy” $\alpha $ by randomly swapping a share ($1-\alpha$) of the predictions for the simulated non-zero observations $z_i^{+}$ with other all-zero predictions, thus keeping the model calibration consistent. For example, for $\alpha = 0.8$ in the 2018 test window, 20\% (255) of the 1,277 prediction samples for simulated non-zero observations are replaced with all-zero predictions, and in turn inserted in the place of random all-zero samples. Second, we vary $\sigma$ to add different levels of random noise $\varepsilon = 50\sigma u\text{, where }u \sim \mathcal{U}(-1, 1) \text{ and } 0 \le \sigma \le 1$ to all simulated non-zero observations before drawing the prediction samples,\footnote{While the cutoff point for the noise multiplier is chosen arbitrarily, the range should be sufficient to produce fairly strong effects given the values seen in our predictions -- based on Figure \ref{fig:fig1} and a median number of fatalities of 5 in the training data. Increasing the multiplier much beyond $50\varepsilon$ continues to decrease the size of the metrics' change when varying the “accuracy”.} i.e. $z_i^{+\prime} = z_i^{+} + \varepsilon$, randomly shifting the center of our predictive distributions away from the actual value.\footnote{We replace any values < 0 with 0 before drawing the samples, to ensure the prediction samples only contain non-negative values.} In both cases we still evaluate against the original $z_i$.

Figure \ref{fig:fig3} shows the simulated CRPS value ranges for the share of non-zero observations across all test windows and across varying accuracies and noise levels in relation to our best model. Except for 2021 and 2022, where outliers in the data strongly impact the non-simulated CRPS and MIS scores, our best model's score falls within the value range of the simulation for all three metrics. As the windows with the lowest and highest share of non-zero \emph{pgms}, we exemplarily discuss the simulation results for 2018 and 2023. Without any noise, CRPS values range from 0.007 to 0.172 and 0.010 to 0.252, respectively, dropping roughly linearly from “perfect” predictions to 0.1 accuracy. This means an average CRPS increase of 0.018 (2018) / 0.027 (2023) per 10\% drop in accuracy. The more noise is added to the data, the smaller the change in CRPS becomes when varying the accuracy, with the changes less than half for the maximum level of noise in our simulation. Since accuracy in contrast to noise has an absolute floor, noise almost completely loses its impact on the score for very low levels of $\alpha$. Given that our best models are on average 0.007 points better than the best benchmark in the performance evaluation, this makes us confident that this means a real improvement in performance over the benchmarks rather than random variation. We observe very similar patterns for the MIS (Figure \ref{fig:figa1}) and also for the IGN (Figure \ref{fig:figa2}), with the caveat of the latter being much more sensitive to noise than CRPS and MIS and the noise curve for "perfect" accuracy appearing logarithmic rather than linear (see equation \ref{eq:2}).

\begin{figure}[H]
    \centering
    \includegraphics[width=1\linewidth]{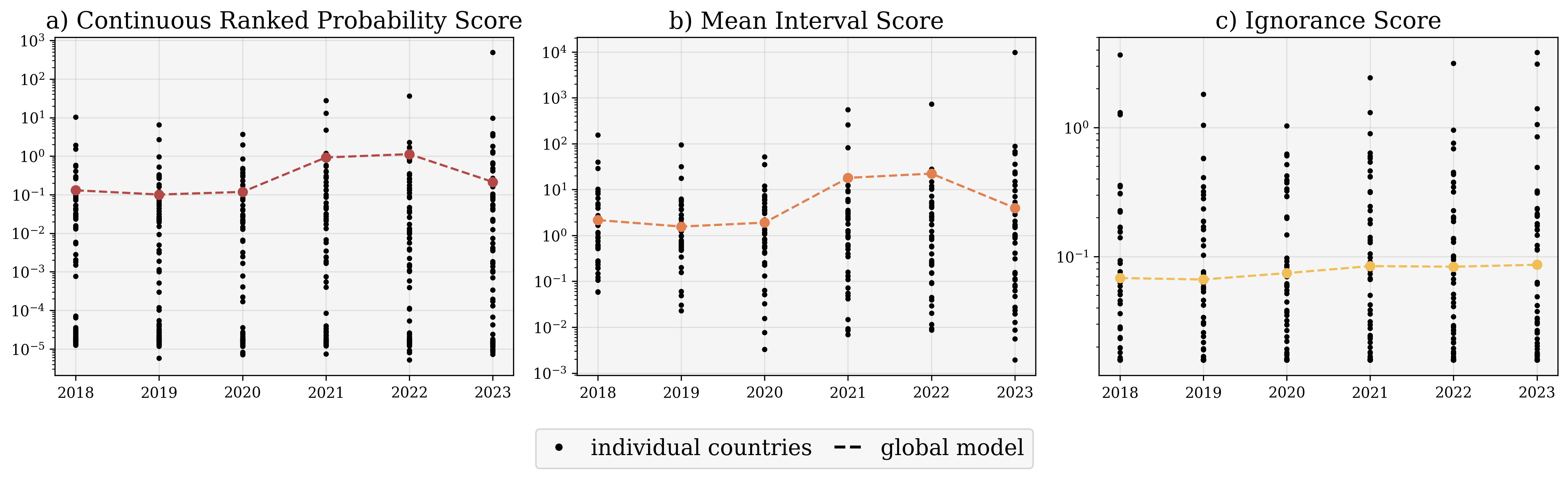}
    \caption{Comparison of model-wide scores with scores based on grid cells grouped by individual countries for the three distributional metrics: a) CRPS, b) MIS, and c) IGN.}
    \label{fig:fig4}
\end{figure}

Finally, having established that the small differences in the evaluation metrics are indeed meaningful, we want to understand where these differences come from. To do so, we match grid cells to countries based on a majority rule and evaluate our global model for each country individually. Country borders for assignment are taken from geoBoundaries \citep{runfola_geoboundaries_2020}. Since this approach means strongly varying levels of zero-inflatedness across the countries, the corresponding scores also vary strongly and cannot be directly compared (Figure \ref{fig:fig4}). Instead, we rank the performance of all models and benchmarks for each country and calculate the average ranks across all scenarios  (Table \ref{tab:table3}). Based on the mean rank across all countries and years, our model only places fifth on the CRPS, fourth on the MIS, and third on the IGN. This can be explained by the fact that roughly half of all countries never experience any violence, for which our model gets heavily penalized in a rank-based evaluation due to its base assumption of a small but non-zero chance of violence anywhere.\footnote{For both the CRPS and the IGN, any non-zero values in our samples mean that predictions will be scored marginally worse. Since the benchmarks generally yield all-zero samples in countries without prior violence, this means our model will be ranked 6th in such cases, while the benchmarks all receive a joint 1st place. To illustrate: Out of 80 countries covered by our model, 32 never see any fatalities across all yearly test windows, with a minimum of 39 no-violence countries in any individual window.} However, when we restrict the analysis to country-year instances with any observed violence -- arguably the cases in which predictive performance is most relevant -- our model achieves the highest average rank compared to all benchmarks across all three distributional metrics. This likely indicates that our models' predictions are relatively better for \emph{pgms} where violence actually occurred.

\begin{table}
\centering
{
    \setlength{\dashlinedash}{2pt}
    \setlength{\dashlinegap}{2pt}
    \setlength{\extrarowheight}{2pt}
    \renewcommand{\ttdefault}{pcr}
    \small\ttfamily
    
    \begin{tabular}{|l!{\vrule width 1.2pt}r|r!{\vrule width 1.2pt}r|r!{\vrule width 1.2pt}r|r|}
    
    \hline
    \rowcolor{gray!30} 
     & \multicolumn{2}{c!{\vrule width 1.2pt}}{\textbf{CRPS}} & \multicolumn{2}{c!{\vrule width 1.2pt}}{\textbf{MIS}} & \multicolumn{2}{c|}{\textbf{IGN}} \\
    \rowcolor{gray!30} 
     \textbf{Model} & \textbf{Mean rank} & \textbf{\shortstack{Mean rank\\non-zero}} & \textbf{Mean rank} & \textbf{\shortstack{Mean rank\\non-zero}} & \textbf{Mean rank} & \textbf{\shortstack{Mean rank\\non-zero}} \\
    \boldhline
    
    \rowcolor{cdark1} Global & \cellcolor{clight1} 3.70 & \textbf{2.43} & \cellcolor{clight1} 1.94 & \textbf{2.58} & \cellcolor{clight1} 2.33 & \textbf{1.55} \\ 
    \rowcolor{cdark1} All-zero & \cellcolor{clight1} 2.08 & 3.24 & \cellcolor{clight1} 1.81 & 2.68 & \cellcolor{clight1} 1.82 & 2.70 \\ 
    \rowcolor{cdark1} Poisson (last) & \cellcolor{clight1} 2.60 & 4.24 & \cellcolor{clight1} 2.24 & 3.51 & \cellcolor{clight1} 2.12 & 3.28 \\ 
    \rowcolor{cdark1} Conflictology & \cellcolor{clight1} 2.38 & 3.55 & \cellcolor{clight1} 2.07 & 2.91 & \cellcolor{clight1} 5.88 & 5.75 \\
    \rowcolor{cdark1} Conf. neighbors & \cellcolor{clight1} 2.32 & 2.83 & \cellcolor{clight1} 2.12 & 2.84 & \cellcolor{clight1} 4.56 & 4.09 \\ 
    \rowcolor{cdark1} Bootstrap 240 & \cellcolor{clight1} 4.85 & 3.91 & \cellcolor{clight1} 1.81 & 2.68 & \cellcolor{clight1} 3.55 & 3.14 \\ 
    \hline

    \end{tabular}
}
\vspace{1em}
\caption{Rank-based evaluation of global model in comparison to benchmarks. Ranks were calculated for all countries based on the underlying grid cells per test window and averaged across countries and test windows. Grid cells were assigned based on the majority of the area within a grid cell covered. Mean rank includes all countries, while mean rank non-zero only includes countries with any fatalities recorded in a given test window. The highest non-zero mean ranks are marked bold.}
\label{tab:table3}
\end{table}

\section{Discussion}
\label{sec:discussion}
We demonstrate that it is possible to produce high-resolution conflict predictions with uncertainty estimates using a modeling architecture built from relatively simple components, which are able to beat a range of heuristic conflict benchmarks and competing approaches in a public prediction challenge. We further find that these performance gains arise from cases where violence occurs, and thus in situations in which prediction tools are likely most valuable to practitioners. With our evaluation, we highlight the need to not simply pursue marginal improvements in evaluation metrics, but to understand how they behave in the context of the given problem. While this is a useful first step towards actionable uncertainty estimates, the prediction intervals we produce are still fairly wide, with further research needed both to sharpen predictions and to reduce the sources of uncertainty.

Building on our approach, data-related avenues for improvement naturally include the use of additional data sources, potentially in combination with multi-level modeling to combine data at different resolutions, while also accounting for the distinction between macro- and micro processes of conflict \citep{balcells_bridging_2014, fritz_predicting_2024}. As our results show, it is possible to combine multiple regional models without significantly losing performance. This opens up avenues to integrate data sources previously discarded due to coverage issues, for example collected by regional organizations such as ECOWAS. Moreover, data cleaning techniques such as outlier correction, applied both to the target and to the features, could reduce data-driven uncertainties. In addition to general improvements through better-suited methods, approaches such as selective classification \citep{el-yaniv_foundations_2010, hullermeier_aleatoric_2021} and out-of-distribution detection \citep{yang_generalized_2024, gruber_sources_2025} could be used on the modeling side to reject predictions where the estimated uncertainty is too high for the predictions to be useful in practice. 

Finally, as little is known about the composition of uncertainty in conflict modeling, more research is needed to understand to what extent uncertainty is caused by data issues and to what extent violence is the consequence of inherently stochastic processes. Much of this question relates back to conflict datasets and the challenges of recording events globally as discussed above. For example, fine-grained local datasets may report many instances of violence not contained in the larger data collection efforts needed for forecasting systems \citep{bazzi_promise_2022}. In combination with differences in inclusion criteria and coding procedures \citep{raleigh_political_2023, oberg_measurement_2025}, this leads to a large number of non-overlapping events among conflict event datasets despite very similar sources \citep{donnay_integrating_2019}, and makes it hard to establish a ground truth for further examination. Addressing this issue would go a long way towards understanding and quantifying the individual components of uncertainty in conflict forecasting, and thus enable a more targeted approach to its reduction.

\section{Replication Code}
The replication code for our modeling pipeline and the visualizations is available on Github at \url{https://github.com/ccew-unibw/uncertaintrees}.

\section{Acknowledgements}
\label{sec:acknowledgements}
This paper is based on a contribution to the VIEWS Prediction Challenge 2023/2024. Financial support for the Prediction Challenge was provided by the German Ministry for Foreign Affairs. For more information on the Prediction Challenge, see \citet{hegre_202324_2025} and 
\url{https://viewsforecasting.org/research/prediction-challenge-2023}.

The Center for Crisis Early Warning is funded by the German Federal Ministry of Defense and the German Federal Foreign Office. The views and opinions expressed in this article are those of the authors and do not necessarily reflect the official policy or position of any agency of the German government.

Generative AI (GPT 5.2) was used to support the creation of the Python code producing the figures in this paper.
\newpage
\bibliography{references}

\begin{thebibliography}{72}
\providecommand{\natexlab}[1]{#1}
\providecommand{\url}[1]{\texttt{#1}}
\expandafter\ifx\csname urlstyle\endcsname\relax
  \providecommand{\doi}[1]{doi: #1}\else
  \providecommand{\doi}{doi: \begingroup \urlstyle{rm}\Url}\fi

\bibitem[Abdar et~al.(2021)Abdar, Pourpanah, Hussain, Rezazadegan, Liu, Ghavamzadeh, Fieguth, Cao, Khosravi, Acharya, Makarenkov, and Nahavandi]{abdar_review_2021}
M.~Abdar, F.~Pourpanah, S.~Hussain, D.~Rezazadegan, L.~Liu, M.~Ghavamzadeh, P.~Fieguth, X.~Cao, A.~Khosravi, U.~R. Acharya, V.~Makarenkov, and S.~Nahavandi.
\newblock A review of uncertainty quantification in deep learning: {Techniques}, applications and challenges.
\newblock \emph{Information Fusion}, 76:\penalty0 243--297, Dec. 2021.
\newblock ISSN 1566-2535.
\newblock \doi{10.1016/j.inffus.2021.05.008}.
\newblock URL \url{https://www.sciencedirect.com/science/article/pii/S1566253521001081}.

\bibitem[Althaus et~al.(2022)Althaus, Peyton, and Shalmon]{althaus_total_2022}
S.~Althaus, B.~Peyton, and D.~Shalmon.
\newblock A {Total} {Error} {Approach} for {Validating} {Event} {Data}.
\newblock \emph{American Behavioral Scientist}, 66\penalty0 (5):\penalty0 603--624, May 2022.
\newblock ISSN 0002-7642.
\newblock \doi{10.1177/00027642211021635}.
\newblock URL \url{https://doi.org/10.1177/00027642211021635}.
\newblock Publisher: SAGE Publications Inc.

\bibitem[Balcells and Justino(2014)]{balcells_bridging_2014}
L.~Balcells and P.~Justino.
\newblock Bridging {Micro} and {Macro} {Approaches} on {Civil} {Wars} and {Political} {Violence}: {Issues}, {Challenges}, and the {Way} {Forward}.
\newblock \emph{Journal of Conflict Resolution}, 58\penalty0 (8):\penalty0 1343--1359, Dec. 2014.
\newblock ISSN 0022-0027.
\newblock \doi{10.1177/0022002714547905}.
\newblock URL \url{https://doi.org/10.1177/0022002714547905}.
\newblock Publisher: SAGE Publications Inc.

\bibitem[Baum and Zhukov(2015)]{baum_filtering_2015}
M.~A. Baum and Y.~M. Zhukov.
\newblock Filtering revolution: {Reporting} bias in international newspaper coverage of the {Libyan} civil war.
\newblock \emph{Journal of Peace Research}, 52\penalty0 (3):\penalty0 384--400, May 2015.
\newblock ISSN 0022-3433, 1460-3578.
\newblock \doi{10.1177/0022343314554791}.
\newblock URL \url{https://journals.sagepub.com/doi/10.1177/0022343314554791}.

\bibitem[Bazzi et~al.(2022)Bazzi, Blair, Blattman, Dube, Gudgeon, and Peck]{bazzi_promise_2022}
S.~Bazzi, R.~A. Blair, C.~Blattman, O.~Dube, M.~Gudgeon, and R.~Peck.
\newblock The {Promise} and {Pitfalls} of {Conflict} {Prediction}: {Evidence} from {Colombia} and {Indonesia}.
\newblock \emph{The Review of Economics and Statistics}, 104\penalty0 (4):\penalty0 764--779, July 2022.
\newblock ISSN 0034-6535, 1530-9142.
\newblock \doi{10.1162/rest_a_01016}.
\newblock URL \url{https://direct.mit.edu/rest/article/104/4/764/97753/The-Promise-and-Pitfalls-of-Conflict-Prediction}.

\bibitem[Bell et~al.(2021)Bell, Spring, Brady, Huang, Squire, Blackwood, Sitter, and Chegini]{bell_xskillscore_2021}
R.~Bell, A.~Spring, R.~Brady, A.~Huang, D.~Squire, Z.~Blackwood, M.~C. Sitter, and T.~Chegini.
\newblock xskillscore: {Metrics} for verifying forecasts, Aug. 2021.
\newblock URL \url{https://zenodo.org/record/5173153}.

\bibitem[Berger(1985)]{berger_statistical_1985}
J.~O. Berger.
\newblock \emph{Statistical {Decision} {Theory} and {Bayesian} {Analysis}}.
\newblock Springer Science \& Business Media, Aug. 1985.
\newblock ISBN 978-0-387-96098-2.
\newblock Google-Books-ID: oY\_x7dE15\_AC.

\bibitem[Bergstra et~al.(2011)Bergstra, Bardenet, Bengio, and Kégl]{bergstra_algorithms_2011}
J.~Bergstra, R.~Bardenet, Y.~Bengio, and B.~Kégl.
\newblock Algorithms for {Hyper}-{Parameter} {Optimization}.
\newblock In \emph{Advances in {Neural} {Information} {Processing} {Systems}}, volume~24. Curran Associates, Inc., 2011.
\newblock URL \url{https://papers.nips.cc/paper_files/paper/2011/hash/86e8f7ab32cfd12577bc2619bc635690-Abstract.html}.

\bibitem[Bergstra et~al.(2013)Bergstra, Yamins, and Cox]{bergstra_making_2013}
J.~Bergstra, D.~Yamins, and D.~Cox.
\newblock Making a {Science} of {Model} {Search}: {Hyperparameter} {Optimization} in {Hundreds} of {Dimensions} for {Vision} {Architectures}.
\newblock \emph{International Conference on Machine Learning}, pages 115--123, 2013.
\newblock ISSN 1938-7228.
\newblock URL \url{https://proceedings.mlr.press/v28/bergstra13.html}.

\bibitem[Blattman and Miguel(2010)]{blattman_civil_2010}
C.~Blattman and E.~Miguel.
\newblock Civil {War}.
\newblock \emph{Journal of Economic Literature}, 48\penalty0 (1):\penalty0 3--57, Mar. 2010.
\newblock ISSN 0022-0515.
\newblock \doi{10.1257/jel.48.1.3}.
\newblock URL \url{https://pubs.aeaweb.org/doi/10.1257/jel.48.1.3}.

\bibitem[Bowlsby et~al.(2020)Bowlsby, Chenoweth, Hendrix, and Moyer]{bowlsby_future_2020}
D.~Bowlsby, E.~Chenoweth, C.~Hendrix, and J.~D. Moyer.
\newblock The {Future} is a {Moving} {Target}: {Predicting} {Political} {Instability}.
\newblock \emph{British Journal of Political Science}, 50\penalty0 (4):\penalty0 1405--1417, Oct. 2020.
\newblock ISSN 0007-1234, 1469-2112.
\newblock \doi{10.1017/S0007123418000443}.
\newblock URL \url{https://www.cambridge.org/core/product/identifier/S0007123418000443/type/journal_article}.

\bibitem[Breiman(2001)]{breiman_random_2001}
L.~Breiman.
\newblock Random {Forests}.
\newblock \emph{Machine Learning}, 45\penalty0 (1):\penalty0 5--32, 2001.
\newblock ISSN 0885-6125.
\newblock \doi{10.1023/A:1010933404324}.

\bibitem[Buhaug and Gleditsch(2008)]{buhaug_contagion_2008}
H.~Buhaug and K.~S. Gleditsch.
\newblock Contagion or {Confusion}? {Why} {Conflicts} {Cluster} in {Space}.
\newblock \emph{International Studies Quarterly}, 52\penalty0 (2):\penalty0 215--233, 2008.
\newblock ISSN 0020-8833.
\newblock \doi{10.1111/j.1468-2478.2008.00499.x}.
\newblock URL \url{https://academic.oup.com/isq/article/52/2/215/1792634}.

\bibitem[Campello et~al.(2013)Campello, Moulavi, and Sander]{campello_density-based_2013}
R.~J. G.~B. Campello, D.~Moulavi, and J.~Sander.
\newblock Density-{Based} {Clustering} {Based} on {Hierarchical} {Density} {Estimates}.
\newblock In J.~Pei, V.~S. Tseng, L.~Cao, H.~Motoda, and G.~Xu, editors, \emph{Advances in {Knowledge} {Discovery} and {Data} {Mining}}, pages 160--172, Berlin, Heidelberg, 2013. Springer.
\newblock ISBN 978-3-642-37456-2.
\newblock \doi{10.1007/978-3-642-37456-2_14}.

\bibitem[Cederman and Weidmann(2017)]{cederman_predicting_2017}
L.-E. Cederman and N.~B. Weidmann.
\newblock Predicting armed conflict: {Time} to adjust our expectations?
\newblock \emph{Science}, 355\penalty0 (6324):\penalty0 474--476, 2017.
\newblock ISSN 1095-9203.
\newblock \doi{10.1126/science.aal4483}.

\bibitem[Cederman et~al.(2013)Cederman, Gleditsch, and Buhaug]{cederman_inequality_2013}
L.-E. Cederman, K.~S. Gleditsch, and H.~Buhaug.
\newblock \emph{Inequality, grievances, and civil war}.
\newblock Cambridge studies in contentious politics. Cambridge University Press, Cambridge, 2013.
\newblock ISBN 978-1-139-08416-1 978-1-107-60304-2 978-1-107-01742-9.
\newblock \doi{10.1017/CBO9781139084161}.

\bibitem[Cevid et~al.(2022)Cevid, Michel, Näf, Bühlmann, and Meinshausen]{cevid_distributional_2022}
D.~Cevid, L.~Michel, J.~Näf, P.~Bühlmann, and N.~Meinshausen.
\newblock Distributional {Random} {Forests}: {Heterogeneity} {Adjustment} and {Multivariate} {Distributional} {Regression}.
\newblock \emph{Journal of Machine Learning Research}, 23\penalty0 (333):\penalty0 1--79, 2022.
\newblock ISSN 1533-7928.
\newblock URL \url{http://jmlr.org/papers/v23/21-0585.html}.

\bibitem[Chadefaux(2017)]{chadefaux_conflict_2017}
T.~Chadefaux.
\newblock Conflict forecasting and its limits.
\newblock \emph{Data Science}, 1\penalty0 (1-2):\penalty0 7--17, 2017.
\newblock ISSN 24518484.
\newblock \doi{10.3233/DS-170002}.

\bibitem[Chadefaux and Schincariol(2025)]{chadefaux_endogenous_2025}
T.~Chadefaux and T.~Schincariol.
\newblock Endogenous conflict and the limits of predictive optimization.
\newblock \emph{EPJ Data Science}, 14\penalty0 (1):\penalty0 82, Nov. 2025.
\newblock ISSN 2193-1127.
\newblock \doi{10.1140/epjds/s13688-025-00599-x}.
\newblock URL \url{https://doi.org/10.1140/epjds/s13688-025-00599-x}.

\bibitem[Chen and Guestrin(2016)]{chen_xgboost_2016}
T.~Chen and C.~Guestrin.
\newblock {XGBoost}: {A} {Scalable} {Tree} {Boosting} {System}, 2016.
\newblock URL \url{https://arxiv.org/pdf/1603.02754.pdf}.
\newblock Place: arXiv.

\bibitem[Cheng et~al.(2023)Cheng, Quilodrán-Casas, Ouala, Farchi, Liu, Tandeo, Fablet, Lucor, Iooss, Brajard, Xiao, Janjic, Ding, Guo, Carrassi, Bocquet, and Arcucci]{cheng_machine_2023}
S.~Cheng, C.~Quilodrán-Casas, S.~Ouala, A.~Farchi, C.~Liu, P.~Tandeo, R.~Fablet, D.~Lucor, B.~Iooss, J.~Brajard, D.~Xiao, T.~Janjic, W.~Ding, Y.~Guo, A.~Carrassi, M.~Bocquet, and R.~Arcucci.
\newblock Machine {Learning} {With} {Data} {Assimilation} and {Uncertainty} {Quantification} for {Dynamical} {Systems}: {A} {Review}.
\newblock \emph{IEEE/CAA Journal of Automatica Sinica}, 10\penalty0 (6):\penalty0 1361--1387, June 2023.
\newblock ISSN 2329-9274.
\newblock \doi{10.1109/JAS.2023.123537}.
\newblock URL \url{https://ieeexplore.ieee.org/abstract/document/10141545}.

\bibitem[Collier(2004)]{collier_greed_2004}
P.~Collier.
\newblock Greed and grievance in civil war.
\newblock \emph{Oxford Economic Papers}, 56\penalty0 (4):\penalty0 563--595, June 2004.
\newblock ISSN 1464-3812.
\newblock \doi{10.1093/oep/gpf064}.
\newblock URL \url{https://academic.oup.com/oep/article-lookup/doi/10.1093/oep/gpf064}.

\bibitem[Collier et~al.(2003)Collier, Elliot, Hegre, Hoeffler, Reynal-Querol, and Sambanis]{collier_breaking_2003}
P.~Collier, V.~L. Elliot, H.~Hegre, A.~Hoeffler, M.~Reynal-Querol, and N.~Sambanis.
\newblock \emph{Breaking the {Conflict} {Trap}: {Civil} {War} and {Development} {Policy}}.
\newblock A {World} {Bank} policy research report. World Bank and Oxford University Press, Washington DC and New York, 2003.
\newblock ISBN 978-0-8213-5481-0.
\newblock \doi{10.1596/978-0-8213-5481-0}.
\newblock Backup Publisher: World Bank.

\bibitem[Collier et~al.(2008)Collier, Hoeffler, and Rohner]{collier_beyond_2008}
P.~Collier, A.~Hoeffler, and D.~Rohner.
\newblock Beyond greed and grievance: feasibility and civil war.
\newblock \emph{Oxford Economic Papers}, 61\penalty0 (1):\penalty0 1--27, Mar. 2008.
\newblock ISSN 0030-7653, 1464-3812.
\newblock \doi{10.1093/oep/gpn029}.
\newblock URL \url{https://academic.oup.com/oep/article-lookup/doi/10.1093/oep/gpn029}.

\bibitem[Croicu and Kreutz(2017)]{croicu_communication_2017}
M.~Croicu and J.~Kreutz.
\newblock Communication {Technology} and {Reports} on {Political} {Violence}: {Cross}-{National} {Evidence} {Using} {African} {Events} {Data}.
\newblock \emph{Political Research Quarterly}, 70\penalty0 (1):\penalty0 19--31, Mar. 2017.
\newblock ISSN 1065-9129, 1938-274X.
\newblock \doi{10.1177/1065912916670272}.
\newblock URL \url{https://journals.sagepub.com/doi/10.1177/1065912916670272}.

\bibitem[Davenport and Ball(2002)]{davenport_views_2002}
C.~Davenport and P.~Ball.
\newblock Views to a {Kill}: {Exploring} the {Implications} of {Source} {Selection} in the {Case} of {Guatemalan} {State} {Terror}, 1977-1995.
\newblock \emph{Journal of Conflict Resolution}, 46\penalty0 (3):\penalty0 427--450, June 2002.
\newblock ISSN 0022-0027, 1552-8766.
\newblock \doi{10.1177/0022002702046003005}.
\newblock URL \url{https://journals.sagepub.com/doi/10.1177/0022002702046003005}.

\bibitem[Davies et~al.(2025)Davies, Pettersson, Sollenberg, and Öberg]{davies_organized_2025}
S.~Davies, T.~Pettersson, M.~Sollenberg, and M.~Öberg.
\newblock Organized violence 1989–2024, and the challenges of identifying civilian victims.
\newblock \emph{Journal of Peace Research}, 62\penalty0 (4):\penalty0 1223--1240, July 2025.
\newblock ISSN 0022-3433, 1460-3578.
\newblock \doi{10.1177/00223433251345636}.
\newblock URL \url{https://journals.sagepub.com/doi/10.1177/00223433251345636}.

\bibitem[Dietrich and Eck(2020)]{dietrich_known_2020}
N.~Dietrich and K.~Eck.
\newblock Known unknowns: media bias in the reporting of political violence.
\newblock \emph{International Interactions}, 46\penalty0 (6):\penalty0 1043--1060, Nov. 2020.
\newblock ISSN 0305-0629, 1547-7444.
\newblock \doi{10.1080/03050629.2020.1814758}.
\newblock URL \url{https://www.tandfonline.com/doi/full/10.1080/03050629.2020.1814758}.

\bibitem[Donnay et~al.(2019)Donnay, Dunford, McGrath, Backer, and Cunningham]{donnay_integrating_2019}
K.~Donnay, E.~T. Dunford, E.~C. McGrath, D.~Backer, and D.~E. Cunningham.
\newblock Integrating {Conflict} {Event} {Data}.
\newblock \emph{Journal of Conflict Resolution}, 63\penalty0 (5):\penalty0 1337--1364, May 2019.
\newblock ISSN 0022-0027, 1552-8766.
\newblock \doi{10.1177/0022002718777050}.
\newblock URL \url{https://journals.sagepub.com/doi/10.1177/0022002718777050}.

\bibitem[Duan et~al.(2019)Duan, Avati, Ding, Thai, Basu, Ng, and Schuler]{duan_ngboost_2019}
T.~Duan, A.~Avati, D.~Y. Ding, K.~K. Thai, S.~Basu, A.~Y. Ng, and A.~Schuler.
\newblock {NGBoost}: {Natural} {Gradient} {Boosting} for {Probabilistic} {Prediction}, Aug. 2019.
\newblock URL \url{http://arxiv.org/pdf/1910.03225}.

\bibitem[Earl et~al.(2004)Earl, Martin, McCarthy, and Soule]{earl_use_2004}
J.~Earl, A.~Martin, J.~D. McCarthy, and S.~A. Soule.
\newblock The {Use} of {Newspaper} {Data} in the {Study} of {Collective} {Action}.
\newblock \emph{Annual Review of Sociology}, 30\penalty0 (1):\penalty0 65--80, Aug. 2004.
\newblock ISSN 0360-0572, 1545-2115.
\newblock \doi{10.1146/annurev.soc.30.012703.110603}.
\newblock URL \url{https://www.annualreviews.org/doi/10.1146/annurev.soc.30.012703.110603}.

\bibitem[El-Yaniv and Wiener(2010)]{el-yaniv_foundations_2010}
R.~El-Yaniv and Y.~Wiener.
\newblock On the {Foundations} of {Noise}-free {Selective} {Classification}.
\newblock \emph{Journal of Machine Learning Research}, 11\penalty0 (53):\penalty0 1605--1641, 2010.
\newblock ISSN 1533-7928.
\newblock URL \url{http://jmlr.org/papers/v11/el-yaniv10a.html}.

\bibitem[Fearon and Laitin(2003)]{fearon_ethnicity_2003}
J.~D. Fearon and D.~D. Laitin.
\newblock Ethnicity, {Insurgency}, and {Civil} {War}.
\newblock \emph{American Political Science Review}, 97\penalty0 (01):\penalty0 75--90, Feb. 2003.
\newblock ISSN 0003-0554, 1537-5943.
\newblock \doi{10.1017/S0003055403000534}.
\newblock URL \url{http://www.journals.cambridge.org/abstract_S0003055403000534}.

\bibitem[Fritz et~al.(2024)Fritz, Dworschak, and Mehrl]{fritz_predicting_2024}
C.~Fritz, C.~Dworschak, and M.~Mehrl.
\newblock Predicting uncertainty in stages: {Using} a semiparametric hierarchical hurdle model for predicting distributions of conflict fatalities, June 2024.
\newblock URL \url{https://viewsforecasting.org/wp-content/uploads/Fritz_VIEWSPredictionChallenge2023.pdf}.

\bibitem[Gawlikowski et~al.(2023)Gawlikowski, Tassi, Ali, Lee, Humt, Feng, Kruspe, Triebel, Jung, Roscher, Shahzad, Yang, Bamler, and Zhu]{gawlikowski_survey_2023}
J.~Gawlikowski, C.~R.~N. Tassi, M.~Ali, J.~Lee, M.~Humt, J.~Feng, A.~Kruspe, R.~Triebel, P.~Jung, R.~Roscher, M.~Shahzad, W.~Yang, R.~Bamler, and X.~X. Zhu.
\newblock A survey of uncertainty in deep neural networks.
\newblock \emph{Artificial Intelligence Review}, 56\penalty0 (1):\penalty0 1513--1589, Oct. 2023.
\newblock ISSN 1573-7462.
\newblock \doi{10.1007/s10462-023-10562-9}.
\newblock URL \url{https://doi.org/10.1007/s10462-023-10562-9}.

\bibitem[Grinsztajn et~al.(2022)Grinsztajn, Oyallon, and Varoquaux]{grinsztajn_why_2022}
L.~Grinsztajn, E.~Oyallon, and G.~Varoquaux.
\newblock Why do tree-based models still outperform deep learning on typical tabular data?
\newblock \emph{Advances in Neural Information Processing Systems}, 35:\penalty0 507--520, Dec. 2022.
\newblock URL \url{https://proceedings.neurips.cc/paper_files/paper/2022/hash/0378c7692da36807bdec87ab043cdadc-Abstract-Datasets_and_Benchmarks.html}.

\bibitem[Gruber et~al.(2025)Gruber, Schenk, Schierholz, Kreuter, and Kauermann]{gruber_sources_2025}
C.~Gruber, P.~O. Schenk, M.~Schierholz, F.~Kreuter, and G.~Kauermann.
\newblock Sources of {Uncertainty} in {Supervised} {Machine} {Learning} -- {A} {Statisticians}' {View}, Jan. 2025.
\newblock URL \url{http://arxiv.org/abs/2305.16703}.
\newblock arXiv:2305.16703 [stat].

\bibitem[Hegre et~al.(2021)Hegre, Nygård, and Landsverk]{hegre_can_2021}
H.~Hegre, H.~M. Nygård, and P.~Landsverk.
\newblock Can {We} {Predict} {Armed} {Conflict}? {How} the {First} 9 {Years} of {Published} {Forecasts} {Stand} {Up} to {Reality}.
\newblock \emph{International Studies Quarterly}, 65\penalty0 (3):\penalty0 660--668, Sept. 2021.
\newblock ISSN 0020-8833, 1468-2478.
\newblock \doi{10.1093/isq/sqaa094}.
\newblock URL \url{https://academic.oup.com/isq/article/65/3/660/6124679}.

\bibitem[Hegre et~al.(2022{\natexlab{a}})Hegre, Akbari, Croicu, Dale, Gåsste, Jansen, Landsverk, Leis, Lindqvist-McGowan, Mueller, Rakhmankulova, Randahl, Rauh, Rød, and Vesco]{hegre_forecasting_2022-1}
H.~Hegre, F.~Akbari, M.~Croicu, J.~Dale, T.~Gåsste, R.~Jansen, P.~Landsverk, M.~Leis, A.~Lindqvist-McGowan, H.~Mueller, M.~Rakhmankulova, D.~Randahl, C.~Rauh, E.~G. Rød, and P.~Vesco.
\newblock Forecasting fatalities, May 2022{\natexlab{a}}.
\newblock URL \url{http://uu.diva-portal.org/smash/record.jsf?dswid=939&pid=diva2%3A1667048}.

\bibitem[Hegre et~al.(2022{\natexlab{b}})Hegre, Lindqvist-McGowan, Vesco, Dale, Croicu, and Randahl]{hegre_forecasting_2022}
H.~Hegre, A.~Lindqvist-McGowan, P.~Vesco, J.~Dale, M.~Croicu, and D.~Randahl.
\newblock Forecasting fatalities in armed conflict: {Forecasts} for {April} 2022–{March} 2025, May 2022{\natexlab{b}}.
\newblock URL \url{http://uu.diva-portal.org/smash/get/diva2:1665945/FULLTEXT01.pdf}.

\bibitem[Hegre et~al.(2023)Hegre, Vesco, Colaresi, and Vestby]{hegre_202324_2023}
H.~Hegre, P.~Vesco, M.~Colaresi, and J.~Vestby.
\newblock The 2023/24 {VIEWS} {Prediction} competition: {Predicting} the number of fatalities in armed conflict, with uncertainty, 2023.
\newblock URL \url{https://viewsforecasting.org/wp-content/uploads/VIEWS_2023.24_Prediction_Competition_Invitation.pdf}.

\bibitem[Hegre et~al.(2025)Hegre, Vesco, Colaresi, Vestby, Timlick, Kazmi, Lindqvist-McGowan, Becker, Binetti, Bodentien, Bohne, Brandt, Chadefaux, Drauz, Dworschak, D’Orazio, Frank, Fritz, Gleditsch, Häffner, Hofer, Klebe, Macis, Malaga, Mehrl, Metternich, Mittermaier, Muchlinski, Mueller, Oswald, Pisano, Randahl, Rauh, Rüter, Schincariol, Seimon, Siletti, Tagliapietra, Thornhill, Vegelius, and Walterskirchen]{hegre_202324_2025}
H.~Hegre, P.~Vesco, M.~Colaresi, J.~Vestby, A.~Timlick, N.~S. Kazmi, A.~Lindqvist-McGowan, F.~Becker, M.~Binetti, T.~Bodentien, T.~Bohne, P.~T. Brandt, T.~Chadefaux, S.~Drauz, C.~Dworschak, V.~D’Orazio, H.~Frank, C.~Fritz, K.~S. Gleditsch, S.~Häffner, M.~Hofer, F.~L. Klebe, L.~Macis, A.~Malaga, M.~Mehrl, N.~W. Metternich, D.~Mittermaier, D.~Muchlinski, H.~Mueller, C.~Oswald, P.~Pisano, D.~Randahl, C.~Rauh, L.~Rüter, T.~Schincariol, B.~Seimon, E.~Siletti, M.~Tagliapietra, C.~Thornhill, J.~Vegelius, and J.~Walterskirchen.
\newblock The 2023/24 {VIEWS} {Prediction} challenge: {Predicting} the number of fatalities in armed conflict, with uncertainty.
\newblock \emph{Journal of Peace Research}, 62\penalty0 (6):\penalty0 2070--2087, Nov. 2025.
\newblock ISSN 0022-3433, 1460-3578.
\newblock \doi{10.1177/00223433241300862}.
\newblock URL \url{https://doi.org/10.1177/00223433241300862}.

\bibitem[Humphreys and Weinstein(2008)]{humphreys_who_2008}
M.~Humphreys and J.~M. Weinstein.
\newblock Who {Fights}? {The} {Determinants} of {Participation} in {Civil} {War}.
\newblock \emph{American Journal of Political Science}, 52\penalty0 (2):\penalty0 436--455, Apr. 2008.
\newblock ISSN 0092-5853, 1540-5907.
\newblock \doi{10.1111/j.1540-5907.2008.00322.x}.
\newblock URL \url{https://onlinelibrary.wiley.com/doi/10.1111/j.1540-5907.2008.00322.x}.

\bibitem[Hutter et~al.(2019)Hutter, Kotthoff, and Vanschoren]{hutter_automated_2019}
F.~Hutter, L.~Kotthoff, and J.~Vanschoren, editors.
\newblock \emph{Automated {Machine} {Learning}: {Methods}, {Systems}, {Challenges}}.
\newblock Springer Nature, 2019.
\newblock \doi{10.1007/978-3-030-05318-5}.
\newblock URL \url{https://library.oapen.org/handle/20.500.12657/23012}.
\newblock Accepted: 2020-03-18 13:36:15.

\bibitem[Hüllermeier and Waegeman(2021)]{hullermeier_aleatoric_2021}
E.~Hüllermeier and W.~Waegeman.
\newblock Aleatoric and epistemic uncertainty in machine learning: an introduction to concepts and methods.
\newblock \emph{Machine Learning}, 110\penalty0 (3):\penalty0 457--506, Mar. 2021.
\newblock ISSN 1573-0565.
\newblock \doi{10.1007/s10994-021-05946-3}.
\newblock URL \url{https://doi.org/10.1007/s10994-021-05946-3}.

\bibitem[Lacina(2006)]{lacina_explaining_2006}
B.~Lacina.
\newblock Explaining the {Severity} of {Civil} {Wars}.
\newblock \emph{Journal of Conflict Resolution}, 50\penalty0 (2):\penalty0 276--289, Apr. 2006.
\newblock ISSN 0022-0027.
\newblock \doi{10.1177/0022002705284828}.
\newblock URL \url{https://doi.org/10.1177/0022002705284828}.
\newblock Publisher: SAGE Publications Inc.

\bibitem[Lampinen and Vehtari(2001)]{lampinen_bayesian_2001}
J.~Lampinen and A.~Vehtari.
\newblock Bayesian approach for neural networks—review and case studies.
\newblock \emph{Neural Networks}, 14\penalty0 (3):\penalty0 257--274, Apr. 2001.
\newblock ISSN 0893-6080.
\newblock \doi{10.1016/S0893-6080(00)00098-8}.
\newblock URL \url{https://www.sciencedirect.com/science/article/pii/S0893608000000988}.

\bibitem[Meinshausen(2006)]{meinshausen_quantile_2006}
N.~Meinshausen.
\newblock Quantile {Regression} {Forests}.
\newblock \emph{Journal of Machine Learning Research}, 7:\penalty0 983--999, 2006.
\newblock URL \url{https://www.jmlr.org/papers/volume7/meinshausen06a/meinshausen06a.pdf}.

\bibitem[Mueller and Rauh(2018)]{mueller_reading_2018}
H.~Mueller and C.~Rauh.
\newblock Reading {Between} the {Lines}: {Prediction} of {Political} {Violence} {Using} {Newspaper} {Text}.
\newblock \emph{American Political Science Review}, 112\penalty0 (2):\penalty0 358--375, May 2018.
\newblock ISSN 0003-0554, 1537-5943.
\newblock \doi{10.1017/S0003055417000570}.
\newblock URL \url{https://www.cambridge.org/core/product/identifier/S0003055417000570/type/journal_article}.

\bibitem[Mueller and Rauh(2022)]{mueller_hard_2022}
H.~Mueller and C.~Rauh.
\newblock The {Hard} {Problem} of {Prediction} for {Conflict} {Prevention}.
\newblock \emph{Journal of the European Economic Association}, 20\penalty0 (6):\penalty0 2440--2467, 2022.
\newblock ISSN 1542-4766.
\newblock \doi{10.1093/jeea/jvac025}.

\bibitem[Mullahy(1986)]{mullahy_specification_1986}
J.~Mullahy.
\newblock Specification and testing of some modified count data models.
\newblock \emph{Journal of Econometrics}, 33\penalty0 (3):\penalty0 341--365, Dec. 1986.
\newblock ISSN 0304-4076.
\newblock \doi{10.1016/0304-4076(86)90002-3}.
\newblock URL \url{https://www.sciencedirect.com/science/article/pii/0304407686900023}.

\bibitem[Murphy et~al.(2024)Murphy, Sharpe, and Huang]{murphy_promise_2024}
M.~Murphy, E.~Sharpe, and K.~Huang.
\newblock The promise of machine learning in violent conflict forecasting.
\newblock \emph{Data \& Policy}, 6:\penalty0 e35, 2024.
\newblock ISSN 2632-3249.
\newblock \doi{10.1017/dap.2024.27}.
\newblock URL \url{https://www.cambridge.org/core/product/identifier/S2632324924000270/type/journal_article}.

\bibitem[Pedregosa et~al.(2011)Pedregosa, Varoquaux, Gramfort, Michel, Thirion, Grisel, Blondel, Prettenhofer, Weiss, Dubourg, Vanderplas, Passos, Cournapeau, Brucher, Perrot, and Duchesnay]{pedregosa_scikit-learn_2011}
F.~Pedregosa, G.~Varoquaux, A.~Gramfort, V.~Michel, B.~Thirion, O.~Grisel, M.~Blondel, P.~Prettenhofer, R.~Weiss, V.~Dubourg, J.~Vanderplas, A.~Passos, D.~Cournapeau, M.~Brucher, M.~Perrot, and É.~Duchesnay.
\newblock Scikit-learn: {Machine} {Learning} in {Python}.
\newblock \emph{Journal of Machine Learning Research}, 12\penalty0 (85):\penalty0 2825--2830, 2011.
\newblock URL \url{https://jmlr.csail.mit.edu/papers/v12/pedregosa11a.html}.

\bibitem[Petropoulos et~al.(2022)Petropoulos, Apiletti, Assimakopoulos, Babai, Barrow, Ben~Taieb, Bergmeir, Bessa, Bijak, Boylan, Browell, Carnevale, Castle, Cirillo, Clements, Cordeiro, Cyrino~Oliveira, De~Baets, Dokumentov, Ellison, Fiszeder, Franses, Frazier, Gilliland, Gönül, Goodwin, Grossi, Grushka-Cockayne, Guidolin, Guidolin, Gunter, Guo, Guseo, Harvey, Hendry, Hollyman, Januschowski, Jeon, Jose, Kang, Koehler, Kolassa, Kourentzes, Leva, Li, Litsiou, Makridakis, Martin, Martinez, Meeran, Modis, Nikolopoulos, Önkal, Paccagnini, Panagiotelis, Panapakidis, Pavía, Pedio, Pedregal, Pinson, Ramos, Rapach, Reade, Rostami-Tabar, Rubaszek, Sermpinis, Shang, Spiliotis, Syntetos, Talagala, Talagala, Tashman, Thomakos, Thorarinsdottir, Todini, Trapero~Arenas, Wang, Winkler, Yusupova, and Ziel]{petropoulos_forecasting_2022}
F.~Petropoulos, D.~Apiletti, V.~Assimakopoulos, M.~Z. Babai, D.~K. Barrow, S.~Ben~Taieb, C.~Bergmeir, R.~J. Bessa, J.~Bijak, J.~E. Boylan, J.~Browell, C.~Carnevale, J.~L. Castle, P.~Cirillo, M.~P. Clements, C.~Cordeiro, F.~L. Cyrino~Oliveira, S.~De~Baets, A.~Dokumentov, J.~Ellison, P.~Fiszeder, P.~H. Franses, D.~T. Frazier, M.~Gilliland, M.~S. Gönül, P.~Goodwin, L.~Grossi, Y.~Grushka-Cockayne, M.~Guidolin, M.~Guidolin, U.~Gunter, X.~Guo, R.~Guseo, N.~Harvey, D.~F. Hendry, R.~Hollyman, T.~Januschowski, J.~Jeon, V.~R.~R. Jose, Y.~Kang, A.~B. Koehler, S.~Kolassa, N.~Kourentzes, S.~Leva, F.~Li, K.~Litsiou, S.~Makridakis, G.~M. Martin, A.~B. Martinez, S.~Meeran, T.~Modis, K.~Nikolopoulos, D.~Önkal, A.~Paccagnini, A.~Panagiotelis, I.~Panapakidis, J.~M. Pavía, M.~Pedio, D.~J. Pedregal, P.~Pinson, P.~Ramos, D.~E. Rapach, J.~J. Reade, B.~Rostami-Tabar, M.~Rubaszek, G.~Sermpinis, H.~L. Shang, E.~Spiliotis, A.~A. Syntetos, P.~D. Talagala, T.~S. Talagala, L.~Tashman, D.~Thomakos, T.~Thorarinsdottir, E.~Todini, J.~R.
  Trapero~Arenas, X.~Wang, R.~L. Winkler, A.~Yusupova, and F.~Ziel.
\newblock Forecasting: theory and practice.
\newblock \emph{International Journal of Forecasting}, 38\penalty0 (3):\penalty0 705--871, July 2022.
\newblock ISSN 0169-2070.
\newblock \doi{10.1016/j.ijforecast.2021.11.001}.
\newblock URL \url{https://www.sciencedirect.com/science/article/pii/S0169207021001758}.

\bibitem[Price and Ball(2015)]{price_selection_2015}
M.~Price and P.~Ball.
\newblock Selection bias and the statistical patterns of mortality in conflict.
\newblock \emph{Statistical Journal of the IAOS}, 31\penalty0 (2):\penalty0 263--272, May 2015.
\newblock ISSN 18747655, 18759254.
\newblock \doi{10.3233/sji-150899}.
\newblock URL \url{https://journals.sagepub.com/doi/full/10.3233/sji-150899}.

\bibitem[Racek et~al.(2025)Racek, Thurner, and Kauermann]{racek_capturing_2025}
D.~Racek, P.~W. Thurner, and G.~Kauermann.
\newblock Capturing the spatio-temporal diffusion effects of armed conflict: {A} nonparametric smoothing approach.
\newblock \emph{Journal of the Royal Statistical Society Series A: Statistics in Society}, page qnaf120, July 2025.
\newblock ISSN 0964-1998.
\newblock \doi{10.1093/jrsssa/qnaf120}.
\newblock URL \url{https://doi.org/10.1093/jrsssa/qnaf120}.

\bibitem[Raleigh et~al.(2023)Raleigh, Kishi, and Linke]{raleigh_political_2023}
C.~Raleigh, R.~Kishi, and A.~Linke.
\newblock Political instability patterns are obscured by conflict dataset scope conditions, sources, and coding choices.
\newblock \emph{Humanities and Social Sciences Communications}, 10\penalty0 (1):\penalty0 1--17, Feb. 2023.
\newblock ISSN 2662-9992.
\newblock \doi{10.1057/s41599-023-01559-4}.
\newblock URL \url{https://www.nature.com/articles/s41599-023-01559-4}.
\newblock Publisher: Palgrave.

\bibitem[Runfola et~al.(2020)Runfola, Anderson, Baier, Crittenden, Dowker, Fuhrig, Goodman, Grimsley, Layko, Melville, Mulder, Oberman, Panganiban, Peck, Seitz, Shea, Slevin, Youngerman, and Hobbs]{runfola_geoboundaries_2020}
D.~Runfola, A.~Anderson, H.~Baier, M.~Crittenden, E.~Dowker, S.~Fuhrig, S.~Goodman, G.~Grimsley, R.~Layko, G.~Melville, M.~Mulder, R.~Oberman, J.~Panganiban, A.~Peck, L.~Seitz, S.~Shea, H.~Slevin, R.~Youngerman, and L.~Hobbs.
\newblock {geoBoundaries}: {A} global database of political administrative boundaries.
\newblock \emph{PLOS ONE}, 15\penalty0 (4):\penalty0 e0231866, Apr. 2020.
\newblock ISSN 1932-6203.
\newblock \doi{10.1371/journal.pone.0231866}.
\newblock URL \url{https://journals.plos.org/plosone/article?id=10.1371/journal.pone.0231866}.
\newblock Publisher: Public Library of Science.

\bibitem[Rød et~al.(2024)Rød, Gåsste, and Hegre]{rod_review_2024}
E.~G. Rød, T.~Gåsste, and H.~Hegre.
\newblock A review and comparison of conflict early warning systems.
\newblock \emph{International Journal of Forecasting}, 40\penalty0 (1):\penalty0 96--112, Jan. 2024.
\newblock ISSN 01692070.
\newblock \doi{10.1016/j.ijforecast.2023.01.001}.
\newblock URL \url{https://linkinghub.elsevier.com/retrieve/pii/S0169207023000018}.

\bibitem[Saito and Rehmsmeier(2015)]{saito_precision-recall_2015}
T.~Saito and M.~Rehmsmeier.
\newblock The precision-recall plot is more informative than the {ROC} plot when evaluating binary classifiers on imbalanced datasets.
\newblock \emph{PLoS ONE}, 10\penalty0 (3):\penalty0 e0118432, 2015.
\newblock ISSN 19326203.
\newblock \doi{10.1371/journal.pone.0118432}.

\bibitem[Schutte and Weidmann(2011)]{schutte_diffusion_2011}
S.~Schutte and N.~B. Weidmann.
\newblock Diffusion patterns of violence in civil wars.
\newblock \emph{Political Geography}, 30\penalty0 (3):\penalty0 143--152, 2011.
\newblock ISSN 0962-6298.
\newblock \doi{10.1016/j.polgeo.2011.03.005}.
\newblock URL \url{https://www.sciencedirect.com/science/article/pii/S0962629811000424}.

\bibitem[Sundberg and Melander(2013)]{sundberg_introducing_2013}
R.~Sundberg and E.~Melander.
\newblock Introducing the {UCDP} {Georeferenced} {Event} {Dataset}.
\newblock \emph{Journal of Peace Research}, 50\penalty0 (4):\penalty0 523--532, 2013.
\newblock ISSN 0022-3433.
\newblock \doi{10.1177/0022343313484347}.

\bibitem[Tollefsen et~al.(2012)Tollefsen, Strand, and Buhaug]{tollefsen_prio-grid_2012}
A.~F. Tollefsen, H.~Strand, and H.~Buhaug.
\newblock {PRIO}-{GRID}: {A} unified spatial data structure.
\newblock \emph{Journal of Peace Research}, 49\penalty0 (2):\penalty0 363--374, Mar. 2012.
\newblock ISSN 0022-3433.
\newblock \doi{10.1177/0022343311431287}.
\newblock URL \url{https://doi.org/10.1177/0022343311431287}.
\newblock Publisher: SAGE Publications Ltd.

\bibitem[Tyralis and Papacharalampous(2024)]{tyralis_review_2024}
H.~Tyralis and G.~Papacharalampous.
\newblock A review of predictive uncertainty estimation with machine learning.
\newblock \emph{Artificial Intelligence Review}, 57\penalty0 (4):\penalty0 94, Mar. 2024.
\newblock ISSN 1573-7462.
\newblock \doi{10.1007/s10462-023-10698-8}.
\newblock URL \url{https://doi.org/10.1007/s10462-023-10698-8}.

\bibitem[Vesco et~al.(2022)Vesco, Hegre, Colaresi, Jansen, Lo, Reisch, and Weidmann]{vesco_united_2022}
P.~Vesco, H.~Hegre, M.~Colaresi, R.~B. Jansen, A.~Lo, G.~Reisch, and N.~B. Weidmann.
\newblock United {They} {Stand}: {Findings} from an {Escalation} {Prediction} {Competition}.
\newblock \emph{International Interactions}, pages 1--37, 2022.
\newblock ISSN 0305-0629.
\newblock \doi{10.1080/03050629.2022.2029856}.

\bibitem[{VIEWS}(2026)]{views_leaderboard_nodate}
{VIEWS}.
\newblock Leaderboard: {Live} evaluation of the forecasts submitted to the 2023/2024 edition of {VIEWS} {Prediction} {Challenge}, 2026.
\newblock URL \url{https://viewsforecasting.org/leaderboard/}.

\bibitem[Ward et~al.(2010)Ward, Greenhill, and Bakke]{ward_perils_2010}
M.~D. Ward, B.~D. Greenhill, and K.~M. Bakke.
\newblock The perils of policy by p-value: {Predicting} civil conflicts.
\newblock \emph{Journal of Peace Research}, 47\penalty0 (4):\penalty0 363--375, July 2010.
\newblock ISSN 0022-3433, 1460-3578.
\newblock \doi{10.1177/0022343309356491}.
\newblock URL \url{https://journals.sagepub.com/doi/10.1177/0022343309356491}.

\bibitem[Weidmann(2015)]{weidmann_accuracy_2015}
N.~B. Weidmann.
\newblock On the {Accuracy} of {Media}-based {Conflict} {Event} {Data}.
\newblock \emph{Journal of Conflict Resolution}, 59\penalty0 (6):\penalty0 1129--1149, Sept. 2015.
\newblock ISSN 0022-0027, 1552-8766.
\newblock \doi{10.1177/0022002714530431}.
\newblock URL \url{https://journals.sagepub.com/doi/10.1177/0022002714530431}.

\bibitem[Weidmann(2016)]{weidmann_closer_2016}
N.~B. Weidmann.
\newblock A {Closer} {Look} at {Reporting} {Bias} in {Conflict} {Event} {Data}.
\newblock \emph{American Journal of Political Science}, 60\penalty0 (1):\penalty0 206--218, 2016.
\newblock ISSN 0092-5853.
\newblock \doi{10.1111/ajps.12196}.

\bibitem[Yang et~al.(2024)Yang, Zhou, Li, and Liu]{yang_generalized_2024}
J.~Yang, K.~Zhou, Y.~Li, and Z.~Liu.
\newblock Generalized {Out}-of-{Distribution} {Detection}: {A} {Survey}.
\newblock \emph{International Journal of Computer Vision}, 132\penalty0 (12):\penalty0 5635--5662, Dec. 2024.
\newblock ISSN 1573-1405.
\newblock \doi{10.1007/s11263-024-02117-4}.
\newblock URL \url{https://doi.org/10.1007/s11263-024-02117-4}.

\bibitem[Öberg and Yilmaz(2025)]{oberg_measurement_2025}
M.~Öberg and M.~C. Yilmaz.
\newblock Measurement issues in conflict event data: {Addressing} some misconceptions about what drives differences between human-coded event datasets.
\newblock \emph{Research \& Politics}, 12\penalty0 (3):\penalty0 20531680251362440, July 2025.
\newblock ISSN 2053-1680, 2053-1680.
\newblock \doi{10.1177/20531680251362440}.
\newblock URL \url{https://journals.sagepub.com/doi/10.1177/20531680251362440}.

\bibitem[Østby(2008)]{ostby_polarization_2008}
G.~Østby.
\newblock Polarization, {Horizontal} {Inequalities} and {Violent} {Civil} {Conflict}.
\newblock \emph{Journal of Peace Research}, 45\penalty0 (2):\penalty0 143--162, Mar. 2008.
\newblock ISSN 0022-3433, 1460-3578.
\newblock \doi{10.1177/0022343307087169}.
\newblock URL \url{https://journals.sagepub.com/doi/10.1177/0022343307087169}.

\end{thebibliography}

\newpage
\appendix
\setcounter{figure}{0}
\renewcommand{\thefigure}{A\arabic{figure}}
\renewcommand{\thesection}{Appendix \Alph{section}} 
\section{Simulation Experiment Results for MIS and IGN}
\begin{figure}[H]
    \centering
    \includegraphics[width=.75\linewidth]{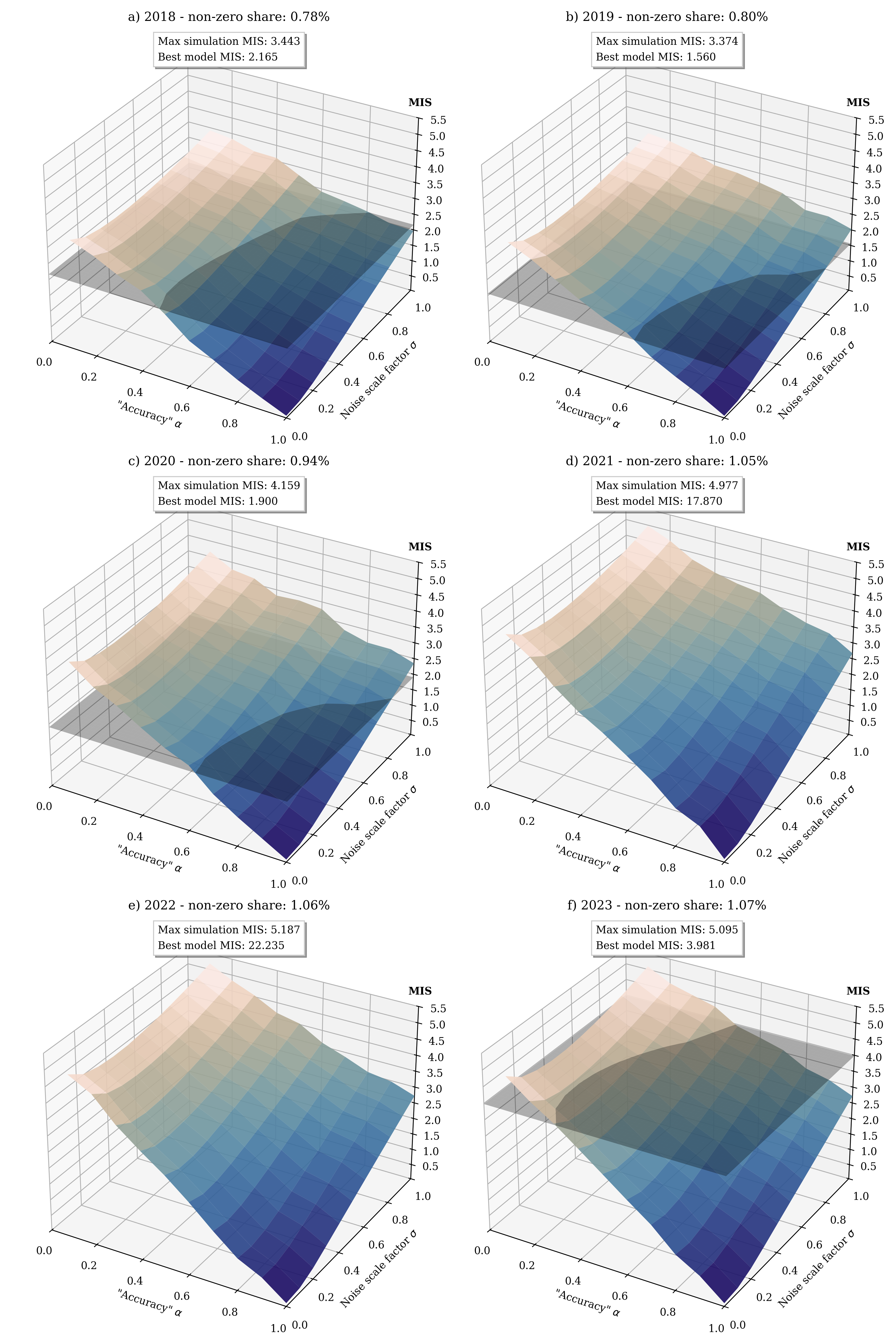}
    \caption{Value ranges of MIS scores with varying noise and “accuracy” based on simulated data. Simulated observations are created with zero-inflatedness matching the six test windows and values drawn from an estimated PDF based on real non-zero observations $< 1,000$ fatalities. Simulated predictions are drawn from a Poisson distribution with mean and variance equal to the corresponding simulated observation. “Accuracy” ($\alpha$) $< 1$ means a share $1-\alpha$ of non-zero predictions was switched with all-zero predictions. The Noise scale factor ($\sigma$) means observations were shifted by $\varepsilon = 50\sigma u$ before creating the samples, with $-1 < u < 1$ drawn from a uniform distribution. The dark planes represents the scores for our best model in the respective test window.}
    \label{fig:figa1}
\end{figure}
\begin{figure}
    \centering
    \includegraphics[width=.75\linewidth]{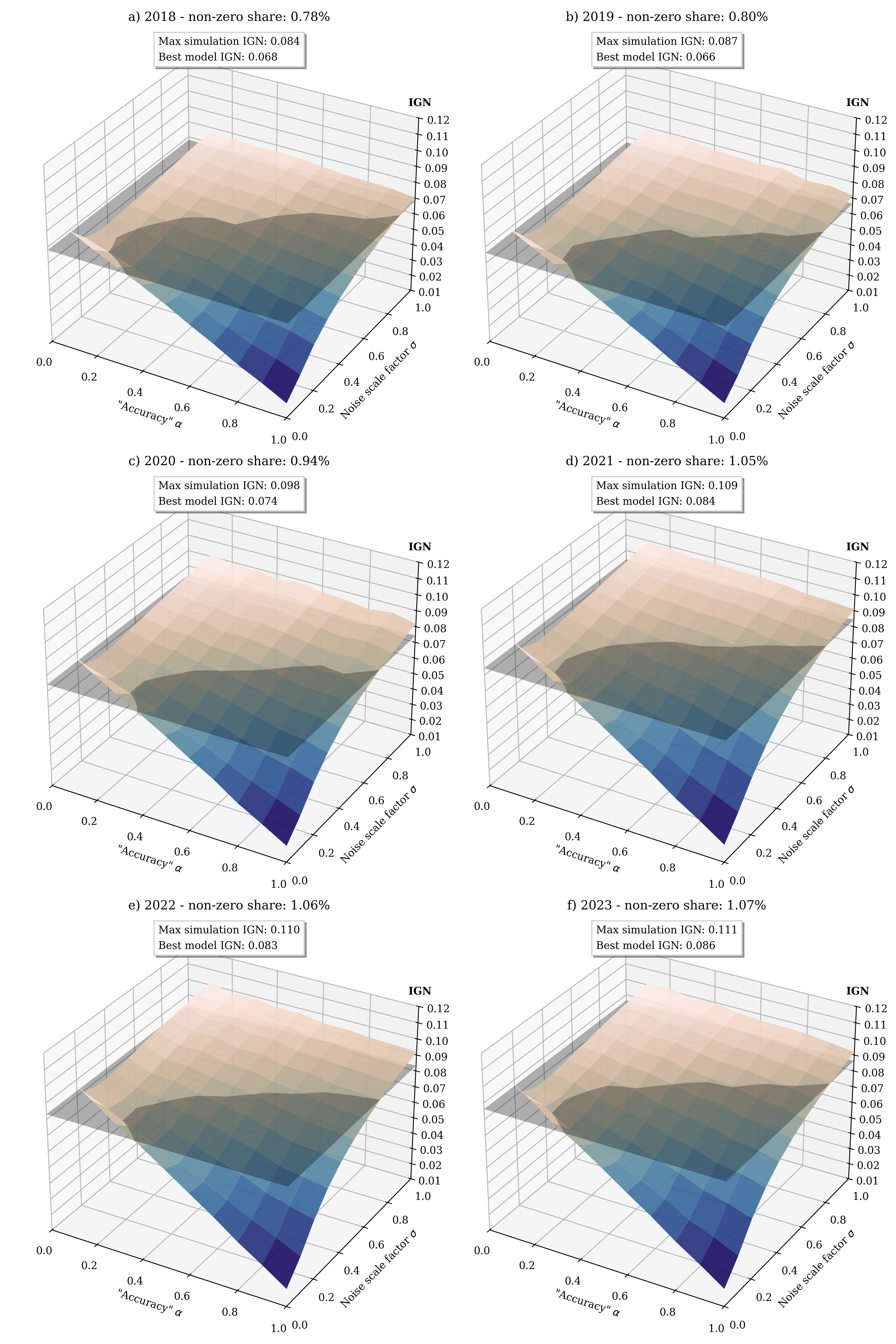}
    \caption{Value ranges of IGN scores with varying noise and “accuracy” based on simulated data. Simulated observations are created with zero-inflatedness matching the six test windows and values drawn from an estimated PDF based on real non-zero observations $< 1,000$ fatalities. Simulated predictions are drawn from a Poisson distribution with mean and variance equal to the corresponding simulated observation. “Accuracy” ($\alpha$) $< 1$ means a share $1-\alpha$ of non-zero predictions was switched with all-zero predictions. The Noise scale factor ($\sigma$) means observations were shifted by $\varepsilon = 50\sigma u$ before creating the samples, with $-1 < u < 1$ drawn from a uniform distribution. The dark planes represents the scores for our best model in the respective test window.}
    \label{fig:figa2}
\end{figure}

\end{document}